\documentclass[%
 aip, pop,
 amsmath,amssymb,
preprint,
]{revtex4-1}
\usepackage{graphicx}
\usepackage{dcolumn}
\usepackage{bm}
\usepackage[utf8]{inputenc}
\usepackage[T1]{fontenc}
\usepackage{amsmath}
\usepackage{natbib}
\usepackage{tikz}

\makeatletter
\def\@email#1#2{%
 \endgroup
 \patchcmd{\titleblock@produce}
  {\frontmatter@RRAPformat}
  {\frontmatter@RRAPformat{\produce@RRAP{*#1\href{mailto:#2}{#2}}}\frontmatter@RRAPformat}
  {}{}
}%
\makeatother
\begin{document}

\newcommand{\red}[1]{{\color{red}#1}}
\newcommand{\iotabar}{\iota\!\!\text{-}}


\title{Integrating Novel Stellarator Single-Stage Optimization Algorithms to Design the Columbia Stellarator Experiment} 



\author{A. Baillod}
\affiliation{Department of Applied Physics and Applied Mathematics, Columbia University, New York, New York 10027, USA}
\author{E. J. Paul}
\affiliation{Department of Applied Physics and Applied Mathematics, Columbia University, New York, New York 10027, USA}
\author{G. Rawlinson}
\affiliation{Department of Applied Physics and Applied Mathematics, Columbia University, New York, New York 10027, USA}
\author{M. Haque}
\affiliation{Department of Applied Physics and Applied Mathematics, Columbia University, New York, New York 10027, USA}
\author{S. W. Freiberger}
\affiliation{Department of Applied Physics and Applied Mathematics, Columbia University, New York, New York 10027, USA}
\author{S. Thapa}
\affiliation{Department of Applied Physics and Applied Mathematics, Columbia University, New York, New York 10027, USA}


\date{\today}

\begin{abstract}

The Columbia Stellarator eXperiment (CSX), currently being designed at Columbia University, aims to test theoretical predictions related to QA plasma behavior, and to pioneer the construction of an optimized stellarator using three-dimensional, non-insulated high-temperature superconducting (NI-HTS) coils. The magnetic configuration is generated by a combination of two circular planar poloidal field (PF) coils and two 3D-shaped interlinked (IL) coils, with the possibility to add windowpane coils to enhance shaping and experimental flexibility. The PF coils and vacuum vessel are repurposed from the former Columbia Non-Neutral Torus (CNT) experiment, while the IL coils will be custom-wound in-house using NI-HTS tapes. 
To obtain a plasma shape that meets the physics objectives with a limited number of coils, novel single-stage optimization techniques are employed, optimizing both the plasma and coils concurrently, in particular targeting a tight aspect ratio QA plasma and minimized strain on the HTS tape. Despite the increased complexity due to the expanded degrees of freedom, these methods successfully identify optimized plasma geometries that can be realized by coils meeting engineering specifications. This paper discusses the derivation of the constraints and objectives specific to CSX, and describe how two recently developed single-stage optimization methodologies are applied to the design of CSX. A set of selected configurations for CSX is then described in detail.

\end{abstract}

\pacs{}

\maketitle 

\section{Introduction}

Stellarators are magnetic confinement devices that rely on 3-dimensional (3D) toroidal magnetic field to confine quasi-neutral plasmas. General 3D magnetic fields do not confine collisionless orbits,\citep{helander_2014} and, in general, stellarators have to be optimized to obtain satisfactory levels of confinement.\citep{helander_2014,landreman_2022,beidler_2021} One way to confine collisionless orbits is to consider quasi-symmetric (QS) magnetic fields,\citep{Boozer_1983,nuhrenberg_1988,helander_2014,rodriguez_2020} i.e. magnetic fields $\mathbf{B}$ that satisfy $B=B(\psi,M\vartheta-N\zeta)$, with $B$ the magnetic field strength, $(\psi,\vartheta,\zeta)$ are the so-called Boozer coordinates,\citep{Boozer_1981} and $(M,N)$ are integers that define the helicity of the symmetry. Quasi-symmetric fields with $M=1,N=0$ are quasi-axisymmetric (QA), while quasi-symmetric fields with $M\neq0,N\neq0$ are quasi-helically (QH) symmetric, and quasi-symmetric fields with $M=0,N=1$ are quasi-poloidally (QP) symmetric.

As of today, only a very limited set of quasi-symmetric experiments have been built. The Helically Symmetric Experiment (HSX),\citep{anderson_1995} at the university of Wisconsin, is an example of a QH field, while the MUSE stellarator \citep{qian_2022,qian_2023} at the Princeton Plasma Physics Laboratory (PPPL) is an example of a QA field. The Chinese First Quasi-Symmetric (CFQS) stellarator, under construction in Southwest Jiaotong University,\citep{kinoshita_2019,isobe_2019} will be another QA experiment. Finally, the National Compact Stellarator Experiment (NCSX), was supposed to be the first QA stellarator,\citep{zarnstorff_2001} but was cancelled before completion.\citep{neilson_2010} Overall, only a few QS experiments exist, and new experiments with different configurations are needed to deepen our understanding of QS fields.

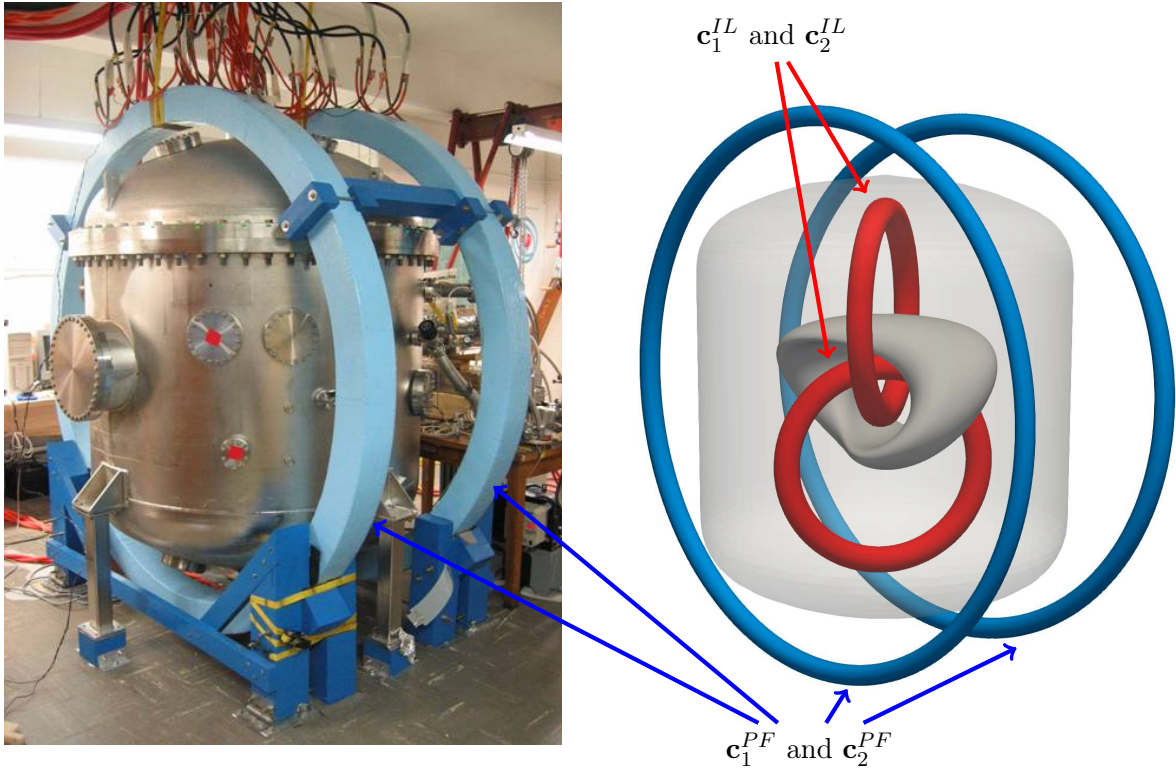
\begin{figure*}
    \centering
    \begin{tikzpicture}
        \node (fig) at (4,-.2) {\includegraphics[width=.65\linewidth]{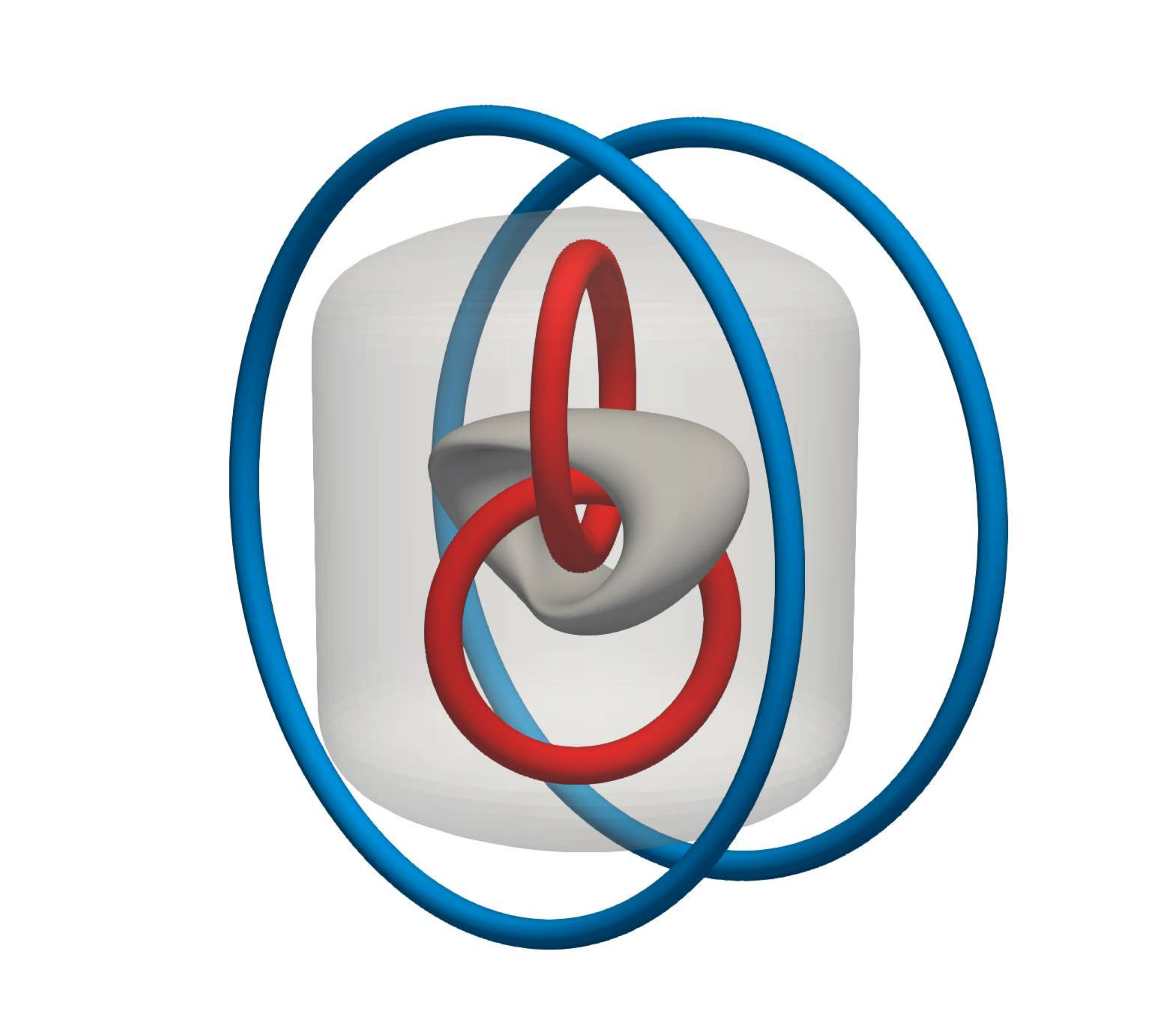}};
        \node (photo) at (-4,0) {\includegraphics[width=0.45\linewidth]{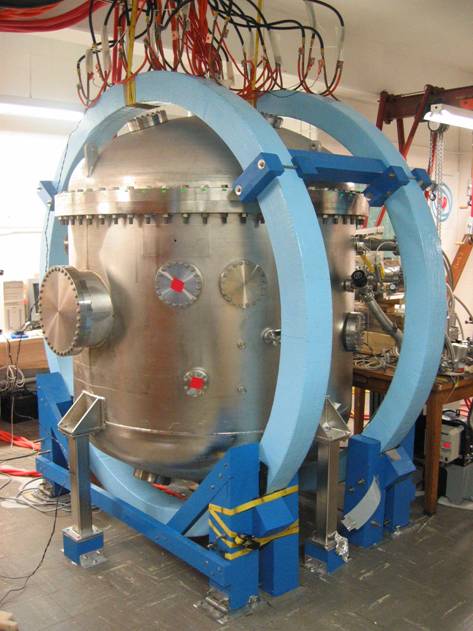}};
        \node (pftext) at (3,-5) {$\mathbf{c}^{PF}_1$ and $\mathbf{c}^{PF}_2$};
        \draw[ultra thick,->,blue] (pftext) -- (3.5,-4.2);
        \draw[ultra thick,->,blue] (pftext) -- (5.75,-3.65);
        \draw[ultra thick,->,blue] (pftext) -- (-2.8,-2);
        \draw[ultra thick,->,blue] (pftext) -- (-1.2,-1.5);
        \node (iltext) at (2.5, 4.5) {$\mathbf{c}^{IL}_1$ and $\mathbf{c}^{IL}_2$};
        \draw[ultra thick,->,red] (iltext) -- (3.75,2.4);
        \draw[ultra thick,->,red] (iltext) -- (3.25,0.2);
    \end{tikzpicture}
    \caption{Picture (left) and sketch (right) of the CNT experiment.}
    \label{fig:cnt_device}
\end{figure*}

The Columbia Non-neutral Torus (CNT) \citep{kremer_2003,pedersen_2004b,pedersen_2006a} is a small scale, compact stellarator at Columbia University. It is composed of a cylindrical vacuum vessel, two circular poloidal field (PF) coils, and two circular, inter-linked (IL) coils (see Figure \ref{fig:cnt_device}). The CNT experiment was designed to study electron and electron-positron plasmas, and relied mostly on a strong electric field for confinement.\citep{pedersen_2002,pedersen_2003,berkery_2006,pedersen_2006b} After many successful campaigns investigating the physics of electron plasmas,\citep{kremer_2006,berkery_2007,berkery_2007d,hahn_2008,marksteiner_2008,hahn_2009} the CNT experiment has been considered to study quasi-neutral plasmas.\citep{hammond_2017,hammond_2017a,hammond_2018} However, as the configuration was not optimized for the confinement of collisionless orbits, the overall confinement properties of the device were poor. 


It has been recently proposed to re-optimize the IL coils of CNT to obtain a QA field, and build the Columbia Stellarator eXperiment (CSX). The external PF coils, as well as the vacuum vessel are kept, while the two IL coils are redesigned and wound using non-planar, non-insulated high temperature superconducting (NI-HTS) technology.\citep{kim_2012, hahn_2011,paz-soldan_2020a} While the CSX experiment is a plasma experiment, it is therefore also a demonstration of a new magnet technology. The coils manufacturing, and the engineering design will be the focus of a separate publication; in this paper, we describe the different optimization algorithms considered to design CSX, and we present some selected configurations.

The CSX will be a two field periods, stellarator symmetric plasma, and is optimized to be QA. In addition, the configuration should have a sufficiently high rotational transform to confine bulk ions, a large volume ($>0.1\ \text{m}^3$), and be filled with magnetic surfaces (\textit{i.e.} without magnetic islands or magnetic field line chaos). The conjunction of the large plasma volume objective, and the constraint of fitting the plasma in the existing CNT vacuum vessel implies a small aspect ratio device, with $A\sim 2$. As the space in the vacuum vessel is limited, only two IL coils are considered. Numerous engineering constraints are imposed on the coils; the IL coils have to fit within the vacuum vessel, and not be too long, as to reduce the amount of HTS tape required. In addition to the usual coil-coil and coil-plasma separation constraints, the coils also must have limited HTS strain.\citep{takayasu_2010,paz-soldan_2020a}

The combination of these different engineering constraints make the CSX optimization challenging, especially with the limited parameter space available for the optimization. Overall, only the shape of one of the IL coils is optimized, as the second one is generated by symmetry. In addition, coils have to be stellarator symmetric\citep{dewar_1998} individually as there are only one coil per field period, further restricting the available coils shape. The limited parameter space available implies that traditional stellarator optimization approaches, \textit{i.e.} the two-stage approach,\citep{nuhrenberg_1988,anderson_1995a} fail for the CSX optimization. Instead, novel optimization approaches \citep{henneberg_2021} are considered, as we discuss now.

The two-stage approach splits the equilibrium optimization calculation from the coil optimization problem. The first step focuses on the magnetic field equilibrium, targeting for example a desired aspect ratio, a target rotational transform, and a QS field.\citep{landreman_2022a} Once a suitable equilibrium is found, the second stage focuses on finding a coil set which is consistent with the optimized plasma boundary.\citep{merkel_1987,drevlak_1998, strickler_2002,strickler_2004,brown_2015,zhu_2017} Note that, in general, no engineering constraints are included in the first stage of this approach, unless proxies for the engineering constraints are derived.\citep{kappel_2023} Therefore, the plasma shape obtained after the first stage can be incompatible with the engineering constraint considered in the second stage, and no suitable coils can be found.\citep{jorge_2023} Furthermore, coil sets found in the second stage of this approach only approximate the optimum magnetic field found in the first stage. Often, the resulting magnetic field produced by the coils is degraded with respect to the configuration found after the first stage.
    
In the case of CSX, the coil parameter space is already limited, reducing the degrees of freedom available during the second stage of the optimization, and strong engineering constraints have to be satisfied. Most optimum plasma shapes found in a stage I optimization would not be compatible with the CSX engineering constraints, as most magnetic fields cannot be generated by a set of two PF and two IL coils. The two-stage approach is therefore not suitable to optimize CSX; instead, novel approaches, called \emph{single-stage optimization methods}, or \emph{combined plasma-coils optimization methods}, have to be considered.\citep{henneberg_2021} With these methods, coils are optimized at the same time as the plasma boundary shape in order to efficiently search for an optimum equilibrium that can be generated by feasible coils. 

To design CSX, two single-stage optimization methods are considered. The first one, thereafter named the \emph{VMEC-based approach}, was recently introduced by \citet{jorge_2023}. It relies on the fixed-boundary magneto hydrodynamic (MHD) code VMEC \citep{hirshman_1986} to evaluate the magnetic equilibrium. It is then coupled with an optimization algorithm using the simsopt Python framework.\citep{landreman_2021} This method successfully found optimized stellarators with the same number of coils as CSX, QA and QH stellarators, and quasi-isodynamic (QI) devices.\citep{jorge_2023} In principle, this approach can also be extended to consider finite-pressure plasma. The second approach considered in this paper, thereafter named the \emph{Boozer surface approach}, does not rely on an equilibrium solver. Instead, it finds an approximate magnetic surface given the magnetic field produced by some coils by solving a partial derivative equation (PDE). The field QA error, rotational transform, as well as the surface aspect ratio and enclosed volume, can then directly be evaluated, without having to solve for an MHD equilibrium. Developed in the past years by \citet{giuliani_2022a,giuliani_2022}, it was successfully coupled with a global optimization algorithm to find more than 300'000 stellarator vacuum magnetic fields close to quasi-symmetry.\citep{jorge2024simplifiedflexiblecoilsstellarators} 


This paper is organized as follows. In section 2, the different objectives and constraints for the CSX optimization are discussed and motivated. The constraints on the coils HTS strain are detailed. In the third section, we discuss how both the VMEC-based approach and the Boozer surface approach can be applied to find a stellarator that satisfies all objectives and constraints described in section 2. We briefly discuss some of their respective advantages and disadvantages. In section 4, multiple configurations obtained with both methods are discussed. Details about the most promising configurations are provided. Finally, section 5 concludes and announces future work on the CSX project.

\section{Objectives and constraints}

The CSX experiment will refurnish the CNT heating system, a $10$kW, $2.45$GHz electron cyclotron heating (ECH) system. Similar plasma parameters as in the CNT device are therefore expected in CSX --- see Table \ref{tab:plasma_parameters}. Note that we expect the electron and ion temperatures to be the same.    

\begin{table}

    \setlength{\tabcolsep}{10pt}
    \centering
    \begin{tabular}{l|r}
         & CNT-like regime   \\
         \hline\hline
       Density          & $10^{17}\ \text{m}^{-3}$  \\
       Temperature & $5$ eV  \\
       Magnetic field strength on axis & $0.1$\ \text{T}  \\
       Volume-averaged plasma $\beta$      & $0.002\%$   \\
       Electron collisionality & $0.25$  \\
       Ion collisionality & $0.17$  \\
    \end{tabular}
    \caption{Expected plasma parameters for the CSX experiment}
    \label{tab:plasma_parameters}
\end{table}

\subsection{Plasma objectives}\label{sec:plasma objectives}
The CSX experiment is designed to study and verify some of the theoretical predictions of neoclassical theory in QA fields. The first objective is then to obtain a magnetic field that is sufficiently close to QA, and that is filled with magnetic surfaces. How close to QA the field has to be is dictated by neoclassical theory; it has been shown \citep{helander_2008} that QS is equivalent to the capability to sustain plasma flow in the direction of symmetry. In recent work, \citet{calvo_2013,calvo_2014,calvo_2015} showed that a magnetic field $\mathbf{B}=\mathbf{B}_0 + \alpha\mathbf{B}_1$ with $\mathbf{B}_0$ a perfectly QA field and $\alpha\mathbf{B}_1$ a non-QA perturbation to $\mathbf{B}_0$, has to satisfy
\begin{equation}
    \alpha < \sqrt{\frac{\rho_i}{|\nabla \text{ln} B_0|^{-1}}}\label{eq.qa_condition}
\end{equation}
to sustain flows, with $\rho_i$ is the ion Larmor radius. Note that other scalings for the flow have been derived by \citet{calvo_2013,calvo_2014,calvo_2015} if the magnetic field does not satisfy Eq.(\ref{eq.qa_condition}); we will not consider these scalings here. The first objective of the CSX optimization is thus to design a magnetic field such that Eq.(\ref{eq.qa_condition}) is satisfied. As it will be detailed later, this is achieved by minimizing the QA error of the magnetic field. Using expected value for CSX parameters, we get $\alpha<0.05-0.1$, \textit{i.e.} the non-QS field of CSX should be smaller than $5-10\%$ of the total magnetic field. 

Note that this criterion is set by the experimental goal to validate flow damping in QA magnetic fields. For larger stellarators, the criterion is usually to have a sufficiently good confinement, which is commonly measured with the effective ripple $\epsilon_{eff}$,\citep{nemov_1999a} while reactors usually require good enough QS to confine energetic particles. In the case of the CSX experiment, lowering the effective ripple is not an objective, as the considered plasma parameters indicate that transport will not be dictated by low-collisionality neoclassical transport. The effective ripple of CSX will nevertheless be improved by several orders of magnitude in comparison to the CNT device following the improvement in quasi-axisymmetry. This is however a feature and not an objective of the optimization.

The second objective for the CSX optimization is to obtain the largest possible plasma volume. Ideally, a plasma volume of a similar order to the CNT plasma volume would be obtained, $V\geq V_c=0.1\ \text{m}^3$. This objective is competing with the QA objective; in general, a larger plasma volume requires a tighter aspect ratio to fit within the vacuum vessel, which comes at the cost of a larger QA error. 

The third objective is to confine most of the bulk ion non-collisional orbits. This is achieved by having a sufficiently good QS field, as discussed above, and by keeping their banana orbit $w_b$ widths small in comparison to the plasma minor radius $a$. For CSX, we target $w_b$ to be smaller than a third of the minor radius. Evaluating the banana width would require to run a particle tracer code for each evaluation of the optimization target function, which could be too computationally expensive. Instead, we derive in Appendix \ref{app.iota_constraint} the minimum rotational transform required to satisfy this constraint, $\iotabar>\iotabar_c=0.27$.


Other important plasma objectives have not been considered in this work. For example, plasma stability to interchange modes could be obtained by enforcing the existence of a magnetic well.\citep{shafranov_1971,greene_1998} Flexibility between different configurations would grant more experimental possibilities.\citep{lee_2022} Finally, one important feature not studied in this work is the sensitivity of the configuration to errors in coils manufacturing and positioning.\citep{wechsung_2022,wechsung_2022a,landreman_2018a} All these will be considered in future work, to further refine candidate configurations for CSX.




\subsection{Coils engineering constraints}
The CSX coils $\mathcal{C}$ are comprised of two PF coils $\mathbf{c}^{PF}_1$ and $\mathbf{c}^{PF}_2$, and two IL coils $\mathbf{c}^{IL}_1$ and $\mathbf{c}^{IL}_2$ (see Figure \ref{fig:cnt_device}). We represent a coil as a closed curve $\mathbf{c}(l)$ in space, and express its position in Cartesian coordinates using Fourier series,
\begin{equation}
    \mathbf{x}(l) = \mathbf{x}_0 + \sum_{k=1}^{N_c} \mathbf{x}^c_{k}\cos(2\pi k l) + \mathbf{x}^s_{k}\sin(2\pi k l),
\end{equation}
with $\mathbf{x}_0$, $\mathbf{x}^c_k$ and $\mathbf{x}^s_k$ the Fourier modes of the curve, $N_c$ the maximum order of the Fourier series, and $l\in[0,1)$ the curve parameter. The effect of the coil finite build is ignored here. The first set of constraints on the IL coils is geometric --- the IL coils have to fit within the vacuum vessel, should not intersect one another, and should not intersect the plasma boundary. In addition, the IL coils should not be too long, in order to minimize the amount of HTS tape needed, and ultimately the price of the device.


The IL coils are wound with NI-HTS tape.\citep{hahn_2011,kim_2012,paz-soldan_2020a} One additional constraint is therefore set by the HTS tape maximum strain. Two kinds of strains are taken into account here; the torsional strain is related to the torsion of the tape, while the binormal curvature strain is associated to the binormal curvature of the tape, also called hard-way bending --- this is the bending of the tape in the same plane as the tape. The sum of both the torsional and binormal curvature strain should not exceed some threshold value set by the tape manufacturer. The critical current and maximum magnetic field in which the HTS tape can operate are, in principle, another constraint - this is however ignored here as the current in each wind of HTS can always be reduced by increasing the number of HTS wind.

\begin{figure}
    \centering
    \begin{tikzpicture}
        \node (fig1) at (-3.75,0) {\includegraphics[height=0.4\linewidth]{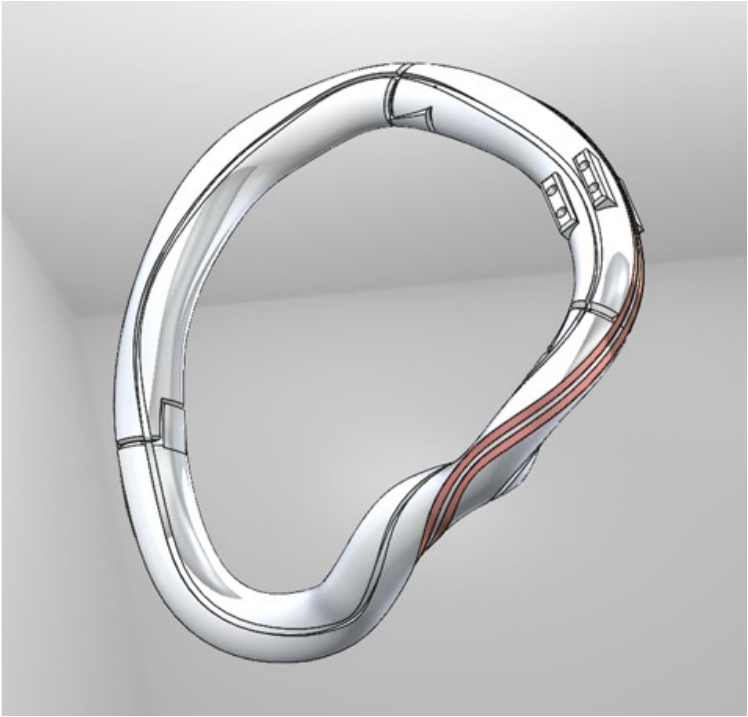}};
        \node (fig2) at ( 3.75,0) {\includegraphics[height=0.3\linewidth]{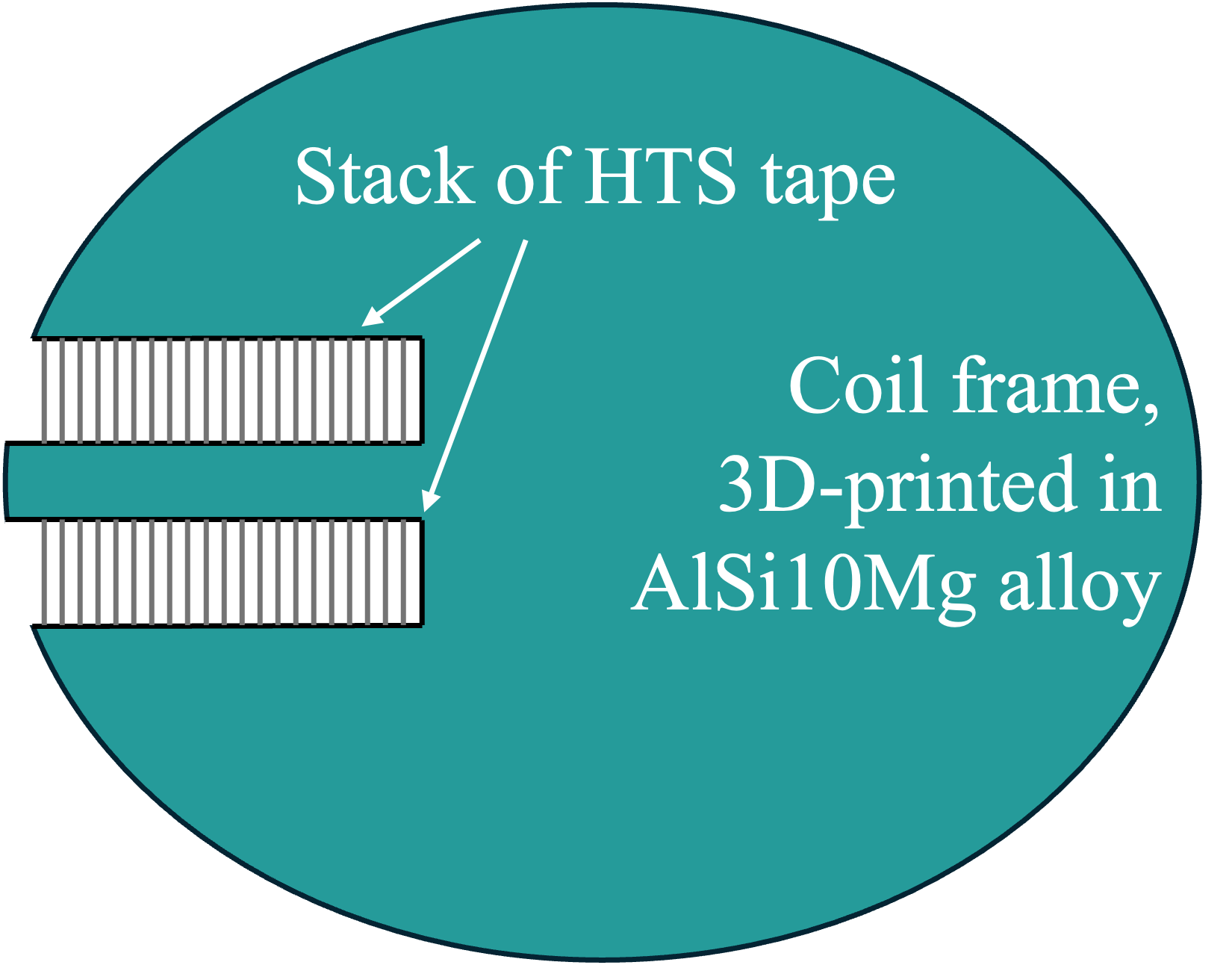}};
    \end{tikzpicture}
    \caption{Sketch of a coil made using NI-HTS tape. Left: 3D rendering of the coil. The red tracks show the position of the HTS tape. Right: cross-section of the 3D-printed coil frame. Two tracks, on the left of the cross-section, form the winding track where the HTS tape will be located.}
    \label{fig:hts_coil}
\end{figure}

\begin{figure}
    \centering
    \includegraphics[width=0.5\linewidth]{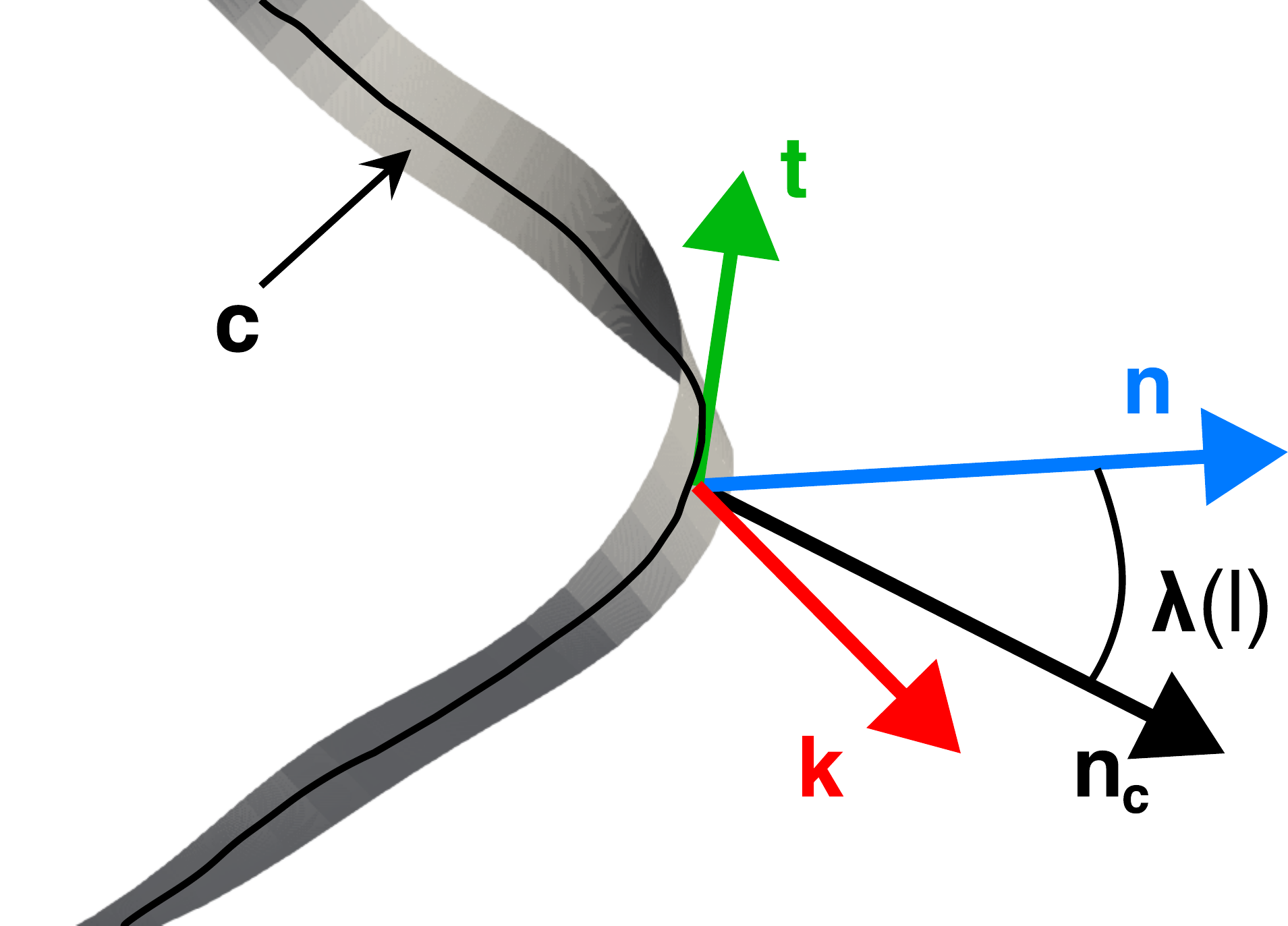}
    \caption{Sketch of the HTS tape orientation and its associated frame. The black arrow is the normal vector associated to the curve centroid frame.}
    \label{fig:winding_angle}
\end{figure}

We describe now some crucial details of the coil design, and how the HTS strain is minimized during the optimization. The frame of the coil is 3D-printed in AlSi10Mg aluminum alloy, with the inclusion of two winding tracks, in which the HTS are wound (see Figure \ref{fig:hts_coil}). Note that the final design of the winding pack is not yet determined --- Figure \ref{fig:hts_coil} is a sketch of a prototype coil. We define the HTS tape frame $(\mathbf{t}, \mathbf{n}, \mathbf{k})$, with $\mathbf{t}$ the curve unit tangent vector, $\mathbf{n}$ the unit vector normal to the tape plane, and $\mathbf{k}$ the unit vector perpendicular to $\mathbf{t}$ and $\mathbf{n}$. The orientation of this frame is fully determined by the angle between the unit vector $\mathbf{n}$ and the curve centroid frame\citep{singh_optimization_2020} normal vector $\mathbf{n}_c$, thereafter named the \emph{winding angle} and denoted by $\lambda(l)$ (see Figure \ref{fig:winding_angle}). We express the winding angle as a Fourier series,
\begin{equation}
    \lambda(l)=\lambda_{c0} + \sum_{n=1}^{N_w} \lambda_{cn}\cos(2\pi n l) + \sum_{n=1}^{N_w}\lambda_{sn}\sin(2\pi n l),
\end{equation}
with $N_w$ the largest Fourier mode order used to describe the winding angle, and $\{\lambda_{cn},\lambda_{sn}\}$ the Fourier coefficients of the winding angle. The additional freedom in choosing the winding angle, \textit{i.e.} the HTS tape orientation, is used to keep the strain on the HTS tape below a critical threshold.\citep{paz-soldan_2020a,huslage2024}

We can now define the binormal curvature and torsional strains of the HTS tape. Here we assume the coils to be wound with a unique channel of HTS tape, and neglect the depth of the channel. Further more detailed studies are required to take these effects into account; in this paper, we design coils with a sufficient margin to the engineering limit such that these higher order corrections to the strain can be safely ignored. The binormal curvature strain \citep{takayasu_2010,paz-soldan_2020a,huslage2024} is defined as
\begin{equation}
    \epsilon_b = \frac{\omega\eta}{2},
\end{equation}
where $w$ is the tape width, and $\eta = \mathbf{k}\cdot\mathbf{t}'/\|\mathbf{c}'\|$ is the binormal curvature of the tape, and where primes denotes derivatives with respect to the arc-length of the curve. The torsional strain \citep{takayasu_2010,paz-soldan_2020a,huslage2024} is defined as
\begin{equation}
    \epsilon_t = \frac{w^2\tau^2}{12},
\end{equation}
with $\tau = \mathbf{k}\cdot\mathbf{n}'/\|\mathbf{c}'\| $ the tape torsion.

All engineering constraints on the coils are gathered in a coil regularization function $f_{reg}(\mathcal{C})$, defined as
\begin{align}
    f_{reg}(\mathcal{C}) &= w_L \max\left(L(\mathbf{c}^{IL}_1)-L_c, 0\right)^2\qquad &\text{Length constraint} \nonumber\\
    &+ w_{cc} \oint_{\mathbf{c}_1^{IL}}\oint_{\mathbf{c}_2^{IL}}\max\left(\|\mathbf{x}_1-\mathbf{x}_2\|-d_{cc}, 0\right)^2dl_1 dl_2 &\text{Coil-coil distance penalty}\nonumber\\
    &+ w_{cs} \oint_{\mathbf{c}_1^{IL}} \iint_\Gamma \max\left(\|\mathbf{x}_1-\mathbf{x}_\Gamma\|-d_{cs}, 0\right)^2 dldS &\text{Coil-plasma distance penalty}\nonumber\\
    &+ w_{cw} \oint_{\mathbf{c}_1^{IL}} \iint_W \max\left(\|\mathbf{x}_1-\mathbf{x}_W\|-d_{cw}), 0\right)^2 dldS &\text{Coil-wall distance penalty}\nonumber\\
    &+ w_{bcs} \oint_{\mathbf{c}_1^{IL}}\max\left[\epsilon_b(\mathbf{c}^{IL}_1, \lambda^{IL}_1)-\epsilon_{b,max}, 0\right]^2 dl &
    \text{HTS binormal curvature strain}\nonumber\\
    &+ w_{tcs} \oint_{\mathbf{c}_1^{IL}}\max\left[\epsilon_t(\mathbf{c}^{IL}_1, \lambda^{IL}_1)-\epsilon_{t,max}, 0\right]^2 dl &
    \text{HTS torsional strain}\nonumber\\
    &+ w_{arc} \sum_{i=1}^{N_q-1}Var(\|\mathbf{x}^i_1-\mathbf{x}_1^{i+1}\|) &\text{Arclength variation} \nonumber\\
    &+ w_{twist} J_{twist}(\mathbf{c}^{IL}_1). & \text{Frame twist penalty} \label{eq.coil_regularization}
\end{align}
Here, the constants $\{w_L, w_{cc}, w_{cs}, w_{cw}, w_{bcs}, w_{tcs}, w_{arc}, w_{twist}\}$ are user-supplied weights. The threshold value $L_c$ sets the maximum curve length for the IL coil, while the distances $d_{cc}=0.06$m, $d_{cs}=0.06$m and $d_{cw}=0.08$m set the minimum distance between the IL coils, the IL coils and the plasma boundary $\Gamma$, and the IL coils and the vacuum vessel $W$ respectively. Here $\mathbf{x}_{1,2}\in \mathbf{c}_{1,2}^{IL}$, $\mathbf{x}_W\in W$ and $\mathbf{x}_\Gamma\in\Gamma$. The function $L(\mathbf{c})$ measures the length of the coil $\mathbf{c}$, $\epsilon_{b,max}$ and $\epsilon_{t,max}$ set the maximum accepted HTS tape binormal curvature strain and torsional strain respectively, and $\epsilon_b(\mathbf{c}, \lambda)$, $\epsilon_t(\mathbf{c}, \lambda)$ measure the binormal curvature and torsional strain respectively. Note that we explicitly denoted the dependence of $\epsilon_b$ and $\epsilon_t$ on the winding angle $\lambda$. All integrals in Eq.(\ref{eq.coil_regularization}) rely on a discrete parameterization of the curves $\mathbf{x}_i$ in a finite number $N_q$ of quadrature points $\mathbf{x}^j_i$. To ensure numerical stability, and penalize curves where quadrature points cluster together, we include as a regularization term the variance of the distance between two consecutive quadrature points along a curve. If $w_{arc}$ is large, quadrature points will be positioned such that the arc-length between each of the consecutive points is constant. In practice, we keep $w_{arc}$ small but non-zero.\citep{wechsung_2022b} The last term of Eq.(\ref{eq.coil_regularization}) penalizes the rotation of the winding frame. Indeed, it is in general harder to wind a coil that is strongly twisted, and this term attempts at avoiding these solutions. More details can be found in the recent publication by \citet{huslage2024}.

\section{Optimization approaches}

\subsection{Single-stage fixed-boundary VMEC optimization \label{sec.vmec_approach}}

The \emph{single-stage fixed-boundary VMEC optimization algorithm}, or \emph{VMEC-based approach}, minimizes the objective function
\begin{align}
    f^{VMEC} = f^{VMEC}_{I}(\Gamma) + w_{coils} \underbrace{\left[f_{reg}(\mathcal{C}) + \iint_\Gamma \left(\frac{\mathbf{B}\cdot\mathbf{\hat n}}{B}\right)^2 dS\right]}_{f_{II}^{VMEC}(\Gamma, \mathcal{C})}, \label{eq.fvmec}
\end{align}
where $f^{VMEC}_I(\Gamma)$ contains all objectives and constraints related to the plasma, \textit{i.e.} what is traditionally optimized in a stage 1 optimization, and $f_{reg}(\mathcal{C})$ contains all the coils regularization terms, \textit{i.e.} the engineering constraints on the coils. The last term of Eq.(\ref{eq.fvmec}), often called the \emph{quadratic flux}, ensures consistency between the coils and the plasma boundary used by VMEC. The sum of the coils' regularization terms and the quadratic flux, hereafter named $f^{VMEC}_{II}$, is the standard objective function commonly used in stage II optimization. 

In this optimization approach, the plasma shape is described using the standard cylindrical coordinates $(R,\phi,Z)$. Using a general poloidal angle $\theta$, we write
\begin{align}
    R(\theta,\phi) &= R_0 +\sum_{m=0}^{m_{pol}}\sum_{n=-n_{tor}}^{n_{tor}} R_{mn}\cos(m\theta-nN_{fp}\phi)\\
    Z(\theta,\phi) &= \sum_{m=0}^{m_{pol}}\sum_{n=-n_{tor}}^{n_{tor}} Z_{mn}\sin(m\theta-nN_{fp}\phi),
\end{align}
where stellarator symmetry is assumed, $N_{fp}$ is the number of field period, and $m_{pol}$ and $n_{tor}$ denote the poloidal and toroidal resolution of the surface.

To obtain a QA magnetic field, we minimized the QA error defined as\citep{helander_2014}
\begin{equation}
    f_{QA} = \sum_{s_j} \left\langle \left[\frac{1}{2\pi B^3}\left(\iotabar M(\nabla B\times \mathbf{B}) \cdot \nabla\psi_t - MG\mathbf{B}\cdot\nabla B\right)\right]^2\right\rangle,
\end{equation}
where $s_j$ is a magnetic surface label, $\langle\cdot\rangle$ denotes a flux surface average, $2\pi G/\mu_0$ is the poloidal current outside the surface $s_j$, with $\mu_0$ the vacuum permeability, and $\psi_t$ is the toroidal flux enclosed by the surface $s_j$.

We construct the plasma target function by writing
\begin{equation}
    f_I^{VMEC}(\Gamma) = f_{QA}(\Gamma) + w_\iotabar \max\left[\iotabar_c-\bar{\iotabar}(\Gamma), 0\right]^2 + w_V \max\left[V_c-V(\Gamma),0\right]^2, \label{eq.fplasma}
\end{equation}
where we explicitly state the dependence on the plasma boundary $\Gamma$, $\bar{\iotabar}$ denotes the rotational transform averaged over the radial profile, and $V$ is the volume enclosed by the plasma boundary $\Gamma$. The scalars $(w_\iotabar,w_V)$ are user-supplied weights that can be adjusted to explore the trade-offs between the different objectives.


Given some plasma boundary, the fixed-boundary VMEC code is used to evaluate the magnetic field within the plasma boundary $\Gamma$, and the term $f^{VMEC}_I(\Gamma)$ is evaluated from the VMEC equilibrium. Derivatives of $f^{VMEC}_I$ with respect to the plasma boundary degrees of freedom are obtained by finite differences. Note that the optimization is performed in vacuum, \textit{i.e.} we assume zero pressure and current when evaluating the fixed-boundary VMEC equilibrium. This is a good approximation for CSX, as the plasma $\beta$ is expected to be negligible --- see Table \ref{tab:plasma_parameters}. The magnetic field generated by the coils is evaluated using the BiotSavart law, and the term $f^{VMEC}_{II}$ can be evaluated from the vacuum field and the coil geometry. The term $f^{VMEC}_{II}$ has analytical expressions, and does not rely on VMEC to be evaluated. Derivatives of $f^{VMEC}_{II}$ are obtained using automatic differentiation algorithms. 

With this optimization approach, the degrees of freedom are then the parameters describing the IL coil shape, the current in the IL and PF coils, the Fourier modes of the IL coil winding angle, $\{\lambda_{cn},\lambda_{sn}\}$, and the parameters describing the plasma shape. All degrees of freedom are optimized at the same time, \textit{i.e.} in the same optimization loop. This causes some challenges, as the number of degrees of freedom can quickly grow as resolution increases. To improve the optimization convergence, we therefore employ a staging approach. During the first optimization, only a limited set of the plasma boundary Fourier harmonics $\{R_{mn},Z_{mn}\}$ with $m,|n|\leq 1$ are optimized. After a first optimum is found, the parameter space is expanded and additional boundary modes are optimized, first with $m,|n|\leq 2$, and then $m,|n|\leq 3,4,\ldots,8$. We use the Broyden–Fletcher–Goldfarb–Shanno (BFGS) algorithm \citep{liu_1989} from the \emph{scipy.optimize} Python package \citep{Virtanen2020} to drive the optimization.

\subsection{Single-stage Boozer surface optimization \label{sec.Boozer_surface_approach}}

The second single-stage optimization algorithm considered is the \emph{single-stage Boozer surface approach}, or \emph{Boozer surface approach}, which we quickly introduce now. More details can be found in the publication by \citet{giuliani_2022}. 


With the Boozer surface approach, only the coils shape and currents are degrees of freedom; the plasma shape, denoted here by $\Gamma_b$, is an output of the calculation. Given the vacuum magnetic field produced by the coils, a magnetic surface is constructed by solving the PDE $\mathbf{r}(\Gamma_b)=0$, with 
\begin{equation}
    \mathbf{r}(\Gamma_b) = G\mathbf{B} - B^2\left(\frac{\partial \mathbf{x}}{\partial \zeta} + \iotabar \frac{\partial \mathbf{x}}{\partial \vartheta}\right), \qquad \mathbf{x}\in\Gamma_b, \label{eq.Boozer_pde}
\end{equation}
and under the constraint that
\begin{equation}
    V(\Gamma_b) - V_c = 0, \label{eq:Boozer_volume}
\end{equation}
with $V_c$ the target volume for the magnetic surface $\Gamma_b$. In this work, we set $V_c=0.1\ \text{m}^3$. Note that when a solution for $\mathbf{r}(\Gamma_b)=0$ is found, the surface $\Gamma_b$ is directly parameterized in Boozer coordinates. 

In practice, a solution to Eq.(\ref{eq.Boozer_pde}) might not exist, for example in the presence of magnetic islands and magnetic field line chaos. Instead, the single-stage Boozer surface optimization algorithm minimizes $\|\mathbf{r}\|^2$ and includes its residual as a coupling term between the coil constraints and the plasma objectives. We then seek to minimize
\begin{equation}
    f^{Boozer}(\mathcal{C}) = f_I^{Boozer}(\Gamma_b(\mathcal{C}))  + f_{reg}(\mathcal{C}) + \frac{1}{2} \iint_{\Gamma_b} \|\mathbf{r}\|^2 dS, \label{eq.target_function_Boozer}
\end{equation}
where $f_I^{Boozer}(\Gamma_b)$ is a term containing all plasma objectives, defined as
\begin{equation}
    f_I^{Boozer}(\Gamma_b) = f^b_{QA}(\Gamma_b) + w_\iotabar \max\left[\iotabar_c- \iotabar(\Gamma_b), 0\right]^2.
\end{equation}
Here $\iotabar(\Gamma_b)$ is the rotational transform on $\Gamma_b$, and the QS error $f^b_{QA}$ is defined as
\begin{equation}
    f^b_{QA}(\Gamma_b) = \left(\frac{\displaystyle\iint_{\Gamma_b} B_{non-QA}^2dS }{\displaystyle\iint_{\Gamma_b} B_{QA}^2 dS}\right)^{1/2}, \label{eq.Boozer_qs_error}
\end{equation}
where the magnetic field is split in its QA and non-QA term, $\mathbf{B} = \mathbf{B}_{QA} + \mathbf{B}_{non-QA}$, by averaging over its toroidal variation on $\Gamma_b$,\citep{giuliani_2022}
\begin{equation}
    \mathbf{B}_{QA}(\Gamma_b) = \frac{\displaystyle\int B\left\| \frac{\partial \mathbf{x}_s}{\partial \zeta}\times \frac{\partial \mathbf{x}_s}{\partial \vartheta} \right\| d\zeta}{\displaystyle\int \left\| \frac{\partial \mathbf{x}_s}{\partial \zeta}\times \frac{\partial \mathbf{x}_s}{\partial \vartheta} \right\| d\zeta}.
\end{equation}
Note that $f_I^{Boozer}$ differs slightly from $f_I^{VMEC}$, but targets the same physical objectives. We take advantage of the magnetic surface being directly parameterized in Boozer coordinates to write a QA objective without requiring further calculations. The rotational transform is only evaluated on the surface $\Gamma_b$. We therefore only constrain the rotational transform on one magnetic surface to be superior to $\iotabar_c$. This is in general not an issue, as solutions approaching QS have a flat rotational transform profile, with low magnetic shear. Finally, there is no need for an additional constraint on the plasma volume $V_c$, as this constraint is enforced when minimizing $\|\mathbf{r}\|^2$. Regarding the coils' regularization term $f_{reg}(\mathcal{C})$, the same terms as in the VMEC-based approach are used. 

The single-stage Boozer surface optimization algorithm is constructed as a nested optimization problem. Given some coils $\mathcal{C}$, the magnetic field $\mathbf{B}$ and the coil regularization term $f_{reg}(\mathcal{C})$ can be evaluated. From the magnetic field, a least-square optimization seeks the surface $\Gamma_b$ that minimizes $\|\mathbf{r}\|^2$. Once $\Gamma_b$ has been constructed, the remaining terms of Eq.(\ref{eq.target_function_Boozer}) are easily evaluated. The outer optimization loop uses then the BFGS algorithm to iterate on the coil set $\mathcal{C}$ to minimize $f^{Boozer}$ .

This algorithm has some advantages over the fixed-boundary VMEC single-stage optimization approach. First, no finite differences are required, as analytical derivatives can be obtained using an adjoint method.\citep{giuliani_2022} Avoiding the use of finite differences helps the algorithm be more robust, and is beneficial to the algorithm performance. Second, minimizing the residuals of $\|\mathbf{r}\|^2$ effectively provides an island healing mechanism --- indeed, a solution to $\mathbf{r}(\Gamma_b)=0$ exists if and only if $\Gamma_b$ is a magnetic surface. Finally, solving the PDE (\ref{eq.Boozer_pde}) in the least squares sense provides robustness to the algorithm. Even if no magnetic surfaces exist, an approximate surface can be found and the optimization does not terminate. There are, however, a few disadvantages to the single-stage Boozer surface optimization algorithm. First, the volume enclosed by $\Gamma_b$ is exactly constrained to be $V_c$ (up to some numerical errors) --- this is a hard constraint, while the VMEC-based approach uses a less constraining inequality constraint on the plasma volume. A second disadvantage, arguably more important than the first, is that this algorithm is only applicable to designing vacuum field, to the opposite of the VMEC-based approach, where finite-$\beta$ optimization are, in principle, possible. This last point is however not an issue in the case of the CSX optimization; indeed, CSX is expected to operate at very low values of $\beta$, which can essentially by approximated by a vacuum field (see Table \ref{tab:plasma_parameters}).

\subsection{Initial guesses}

Different initial guesses are considered to optimize the CSX configuration. To compare each initial guess, we compute the Boozer surface $\Gamma_b$ solving Eqs.(\ref{eq.Boozer_pde})-(\ref{eq:Boozer_volume}) in the least of squares sense. The first two initial guesses considered are some of the CNT device configurations. The CNT device was designed such that the angle between the circular IL coils could be changed to either 32, 39 or 44 degrees, thereby allowing the exploration of three different configurations, thereafter named CNT32, CNT39 and CNT44 respectively. The CNT44 configuration is not considered as an initial guess, as its plasma volume is smaller than the target volume for CSX. A third configuration used as initial guess, thereafter named the \emph{jorge\_QA}, was recently published by \citet{jorge_2023}. The jorge\_QA was obtained using the VMEC-based approach. The last configuration considered as initial guess is obtained from the Compact Stellarator with Simple Coils (CSSC), recently published by \citet{yu_2022}. This configuration however does not share the same PF coils as CNT, and does not fit within the vacuum vessel. We therefore substitute the original CSSC PF coils for the CNT PF coils geometry and perform a first optimization of the IL coils to produce a new configuration, thereafter named the rescaled CSSC configuration (rCSSC).

\begin{figure}
    \centering
    \begin{tikzpicture}
        \node (f1) at (-4, 4) {\includegraphics[width=0.55\linewidth]{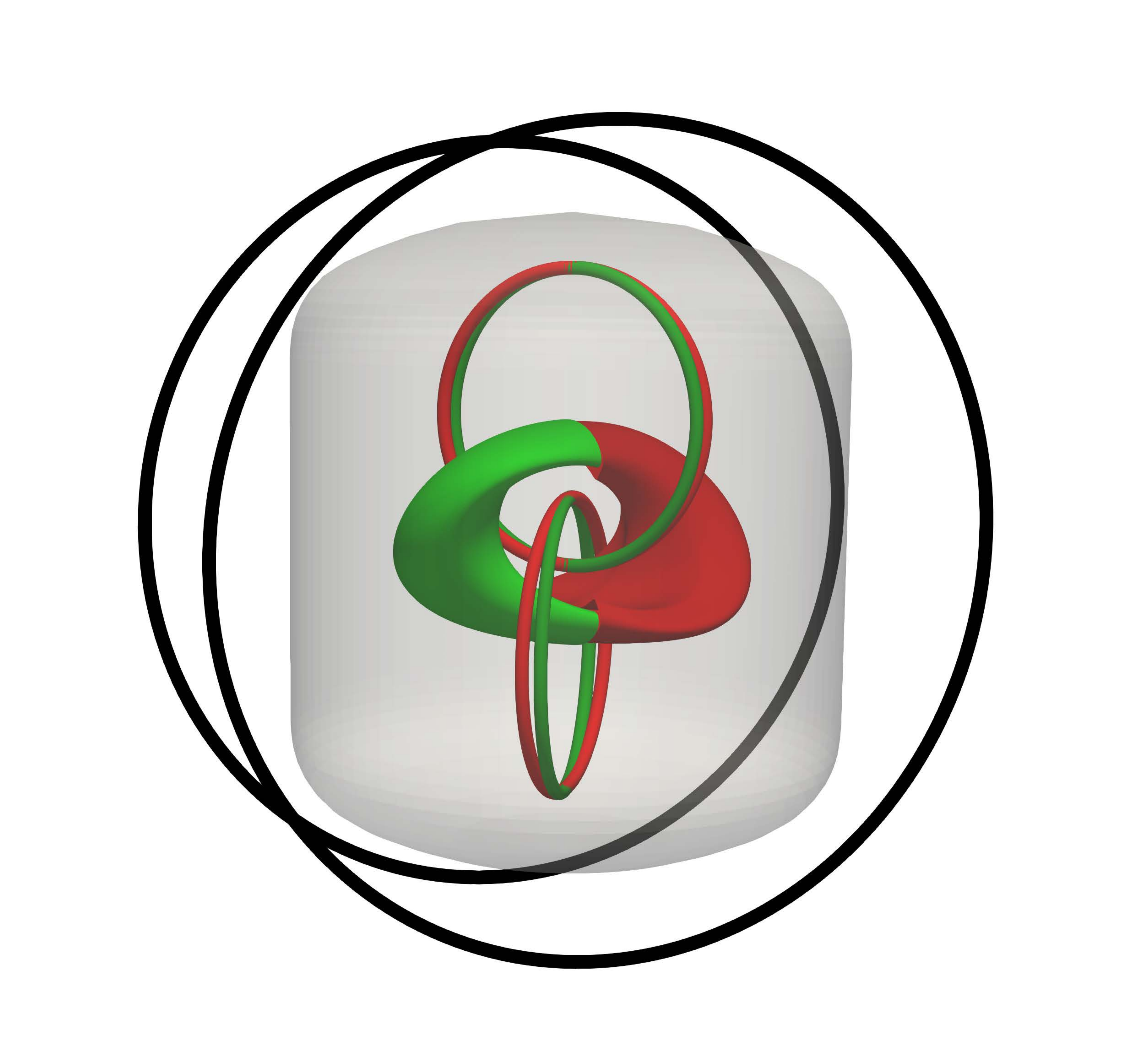}};
        \node (f2) at ( 4, 4) {\includegraphics[width=0.55\linewidth]{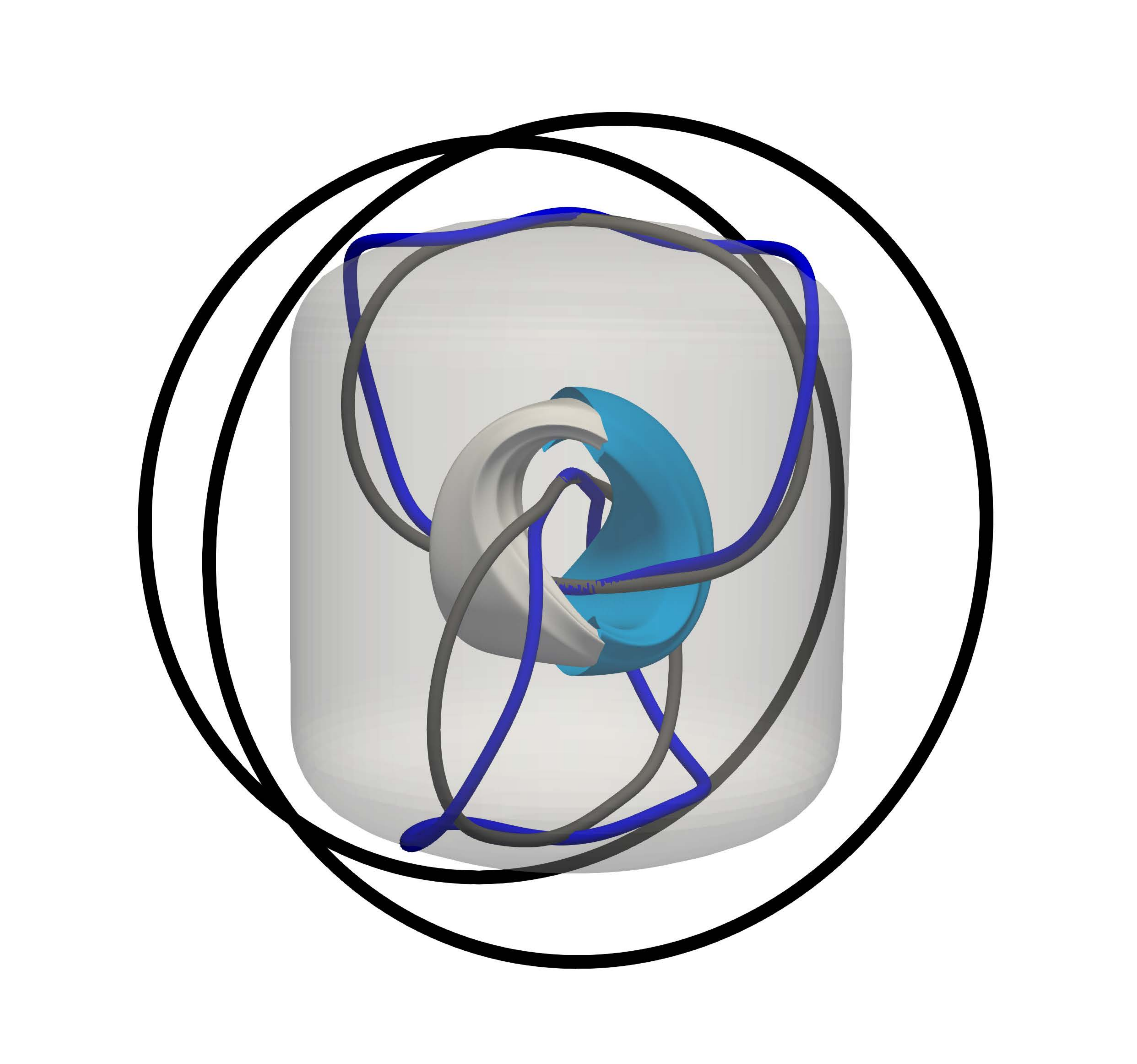}};
        \node (f5) at (-3.85, -3.75) {\includegraphics[height=7.5cm]{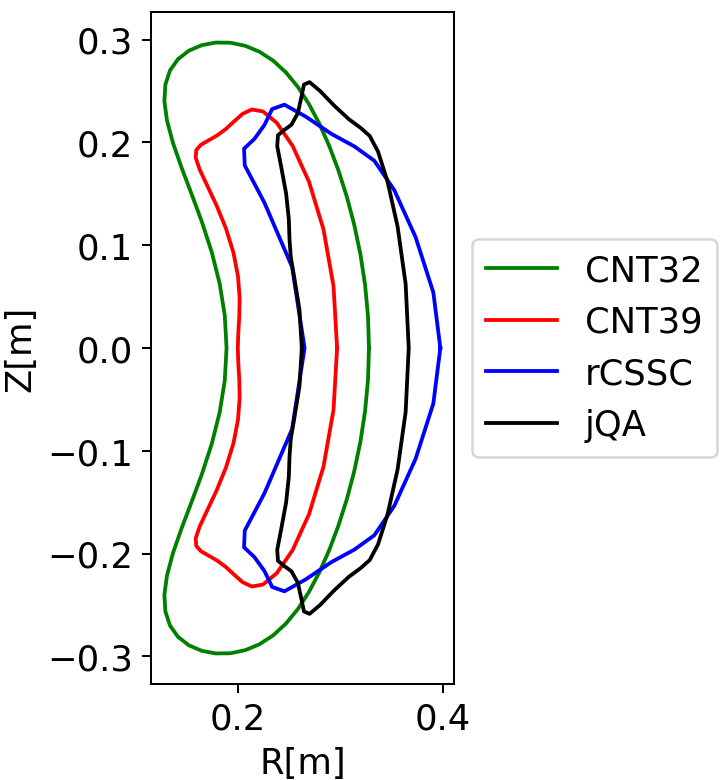}};
        \node (f6) at ( 3.6, -3.4) {\includegraphics[height=6cm]{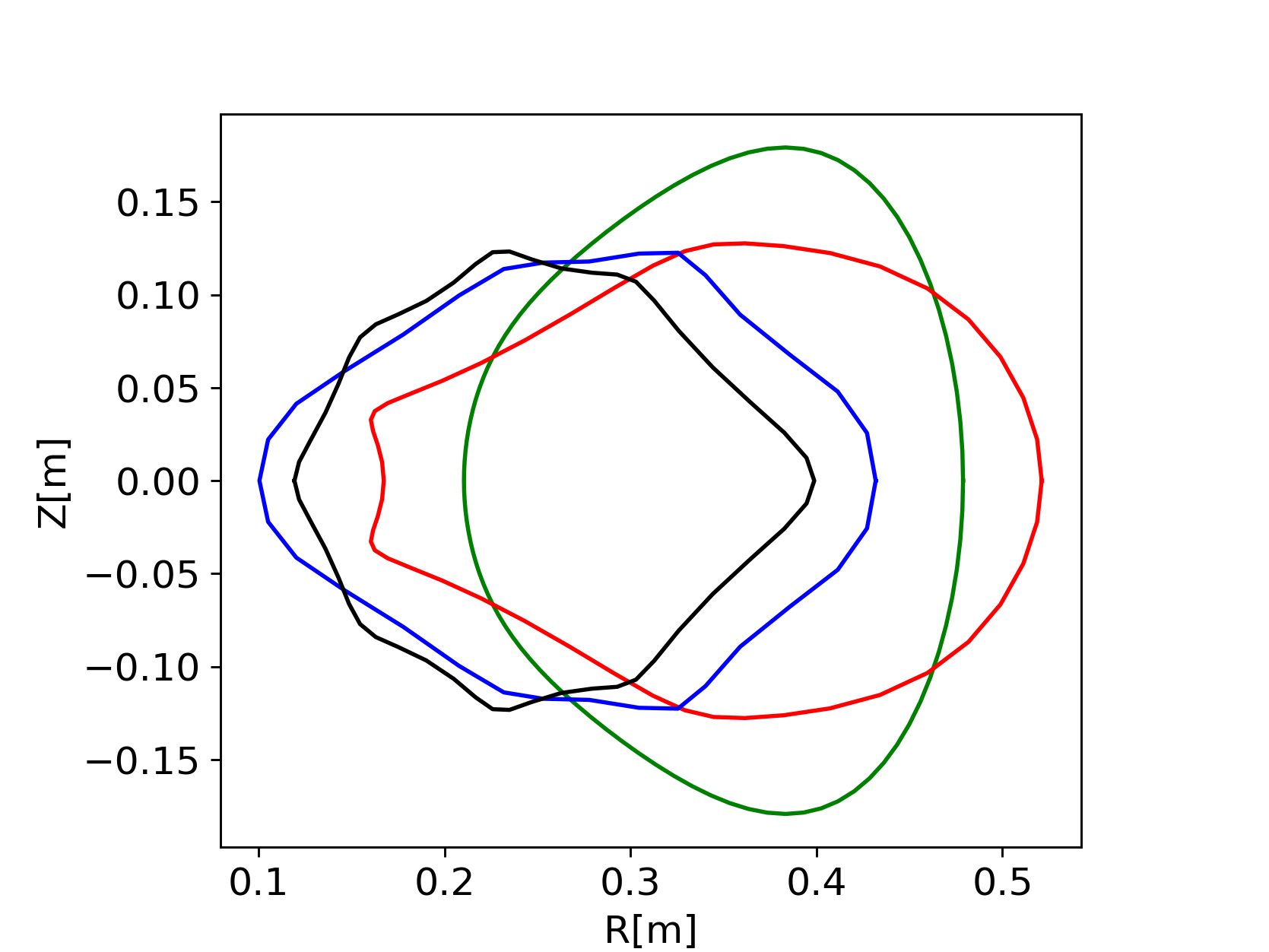}};
    \end{tikzpicture}
    \label{fig:inital_guesses_geometry}
\end{figure}

Some key figures of merit are listed in Table \ref{tab:initial guesses}. Initial guesses have magnetic surfaces extending further out from $\Gamma_b$, excepted for the jorge\_QA configuration. This is indicated as the volume enclosed by the last closed flux surface (LCFS) in table \ref{tab:initial guesses}. The CNT32 and CNT39 initial guesses both have large symmetry breaking modes, as CNT was not designed to be QA. The jorge\_QA and rCSSC both have smaller QA errors. The jorge\_QA volume and rotational transform are however smaller than the targeted values for CSX. The coils' geometry and plasma shapes are shown in Figure \ref{fig:inital_guesses_geometry}. All coils satisfy the engineering constraints, excepted the rCSSC coils, as they do not fit within the vacuum vessel. Starting from the CNT32, CNT39, or the jorge\_QA initial guess, the challenge is to improve the plasma figures of merits without breaking any of the engineering constraints. The opposite is true when starting from the rCSSC initial guess, as the challenge is now to change the coils such that they satisfy the engineering constraint, without impacting the plasma figures of merit.  


\begin{table}

    \setlength{\tabcolsep}{10pt}
    \centering
    \begin{tabular}{l|rrr}
                  &  $\iotabar(\Gamma_b)$ & LCFS Volume $[\text{m}^3]$ & $f^b_{QA}(\Gamma_b)$ \\
        \hline\hline
        CNT32     &  $0.21$               & $0.26$                     & $31.9\%$             \\
        CNT39     &  $0.37$               & $0.11$                     & $29.5\%$             \\
        jorge\_QA &  $0.24$               & $0.08$                     & $7.42\%$             \\
        rCSSC     &  $0.27$               & $0.20$                     & $5.43\%$             
    \end{tabular}
    \caption{Figures of merit for the considered initial guesses.}
    \label{tab:initial guesses}
\end{table}


In summary, the CNT32 and CNT39 configuration both have a relatively large plasma volume, at the expense of a large QS error, while the jorge\_QA configuration has a small QS error, but a small plasma volume. The rCSSC initial guess has the target plasma volume and QS error, but does not satisfy the target rotational transform and the engineering constraints on the coils. The optimization goal is to bridge the gap between these initial guesses, and obtain a large plasma volume, with a small QS error, and coils that satisfy the engineering constraints.

\section{Results}
\subsection{Exploration of parameter space \label{sec:best_results}}
We discuss now a set of optimum configurations found for different values of $L_c=\{4.5,4.75,5.0\}$m, starting from different initial guesses. Results obtained with both the Boozer surface approach and the VMEC-based approach are presented.

To compare the two algorithms, we proceed as follows. Given the output of the VMEC approach, we construct a Boozer surface $\Gamma_b$ by minimizing $\mathbf{r}^2(\Gamma)=0$, given by Eq.(\ref{eq.Boozer_pde}). In general, this surface is slightly different from the VMEC boundary, as the last term of the objective function (\ref{eq.fvmec}) is not exactly zero, and the Boozer surface approximates better the magnetic surfaces produced by the coils than the VMEC boundary. We then evaluate the QS residual, rotational transform, and enclosed plasma volume, on $\Gamma_b$. In the case of the Boozer surface algorithm, the surface $\Gamma_b$ is the output of the optimization and the relevant figures of merit can directly be evaluated.

\begin{figure}
    \centering
    \includegraphics[width=0.75\linewidth]{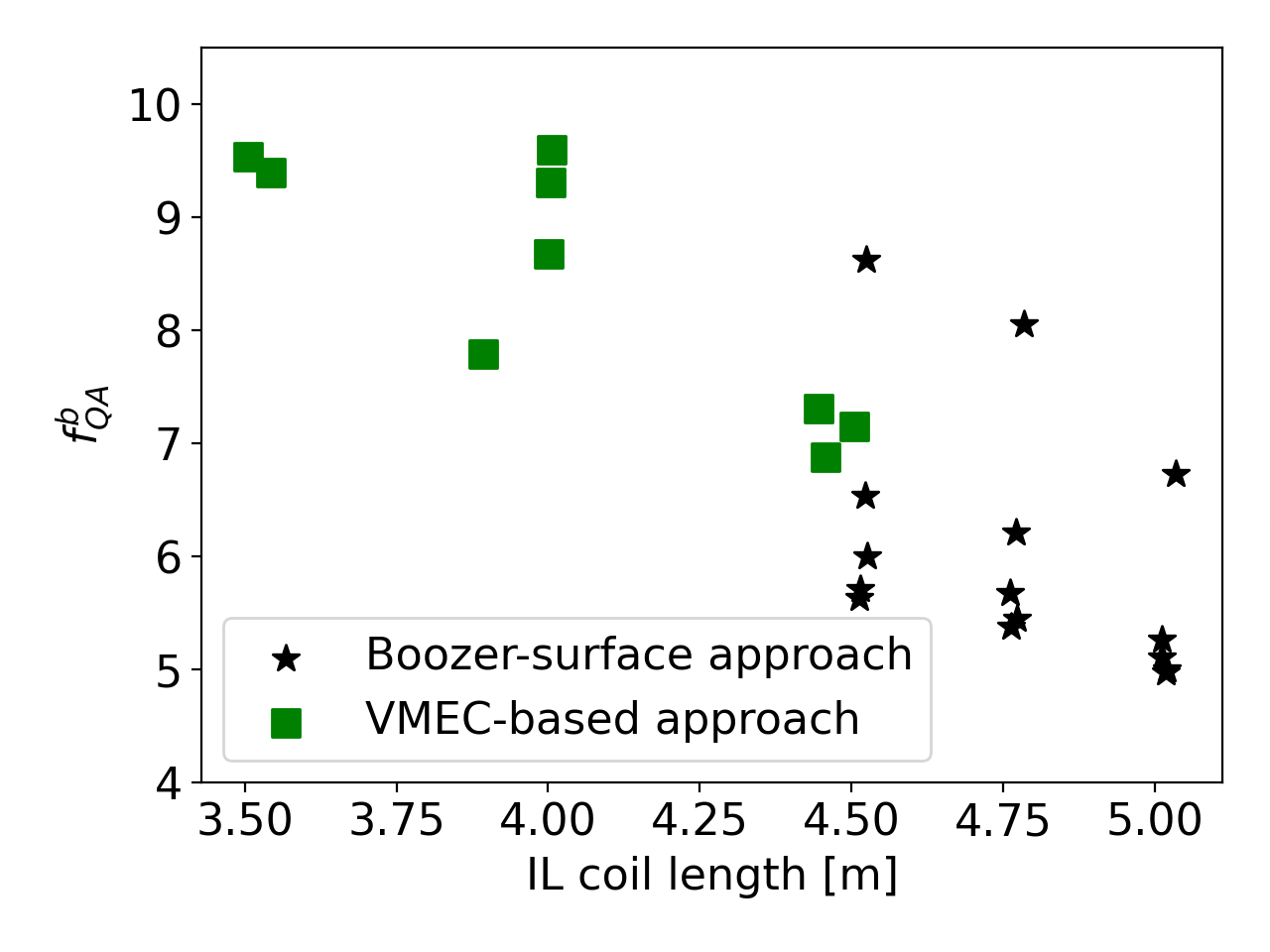}
    \caption{QS error as evaluated by Eq.(\ref{eq.Boozer_qs_error}) as a function of the IL coil length for a set of optimum configurations obtained with the VMEC-based approach (green squares) and the Boozer surface approach (black stars).}
    \label{fig:vmec_vs_Boozer_surface_configuration_summary}
\end{figure}

\begin{figure}
    \centering
    \includegraphics[width=0.8\linewidth]{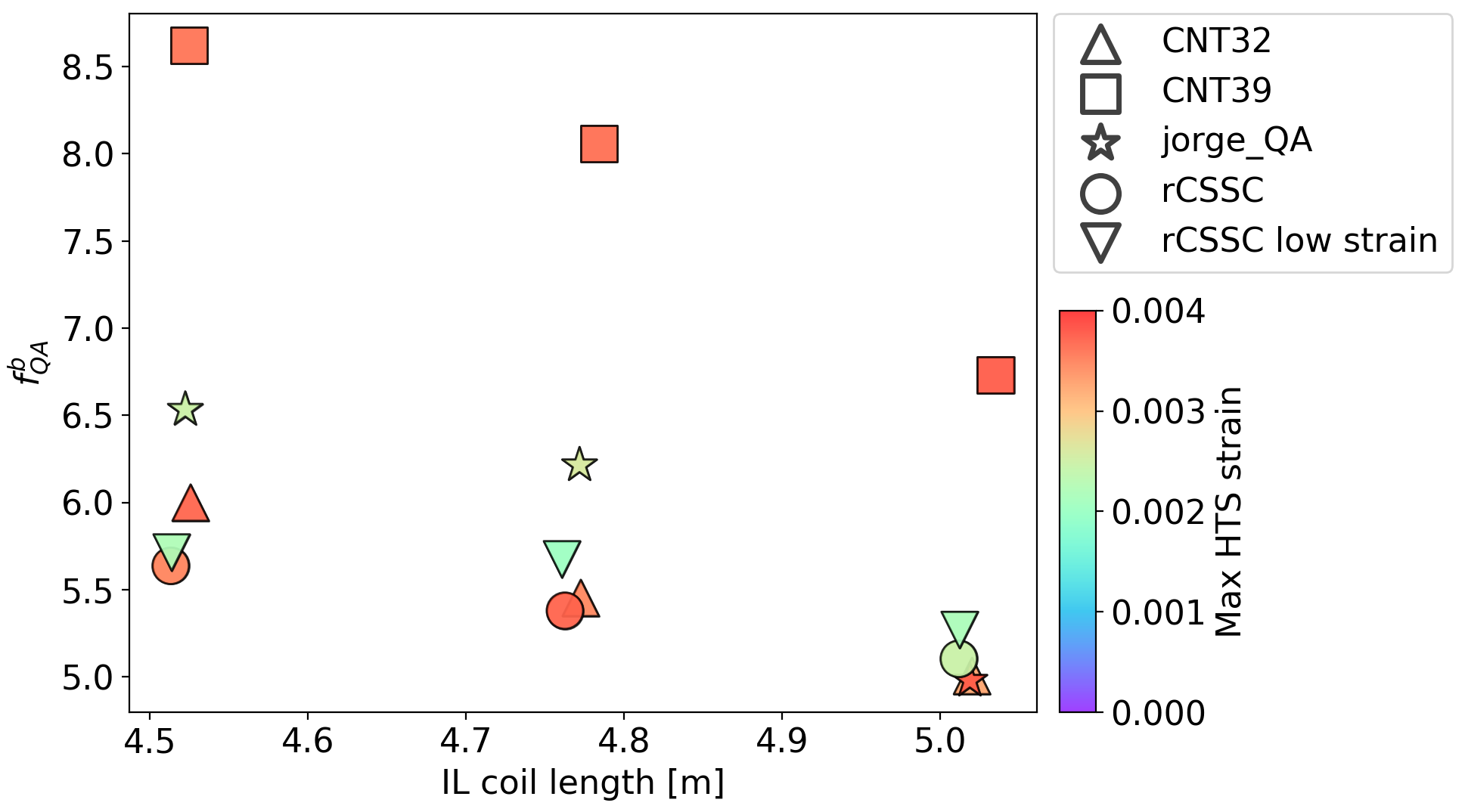}
    \caption{Summary of optimum configurations obtained with the Boozer surface approach. QS error as evaluated by Eq.(\ref{eq.Boozer_qs_error}) as a function of the IL coil length. The inverted triangles are configurations obtained using the rCSSC initial guess, but where lower HTS strain were sought. Colors indicate the maximum HTS strain along the curve.}
    \label{fig:optimum_summary}
\end{figure}

In Figure \ref{fig:vmec_vs_Boozer_surface_configuration_summary}, the QS error is plotted as a function of the IL coil length. Both configurations obtained with the VMEC-based approach (green squares) and those obtained with the Boozer surface approach (black stars) are plotted. Details about the dependence of the optima founds with the Boozer surface approach on the initial guess is shown in Figure \ref{fig:optimum_summary}.


Notice that the results obtained with the VMEC-based approach have consistently larger QS error than the results obtained with the Boozer surface approach. We provide four arguments below explaining why the Boozer surface approach perform better.


First, notice that all configurations found with the VMEC-based approach have shorter IL coils than the configurations found with the Boozer surface approach. In practice, the VMEC-based approach requires lots of fine-tuning of the user-supplied weights --- choosing the wrong weights often lead to solutions where no magnetic surfaces exist. This fine-tuning was found to be increasingly difficult for longer IL coils, and no solutions were found using the VMEC-based approach with coils longer than $4.5$m. The Boozer-surface approach, on the other hand, does not suffer from this limitation, and results with longer coils were routinely found.

Second, VMEC fails to find an equilibrium in some regions of parameter space. This is especially true as CSX has a tight aspect ratio, which can impact negatively VMEC capability to converge towards an equilibrium. In practice, the objective function returns a large value when VMEC does not find an equilibrium, implying that entire regions of parameter space are avoided by the optimizer. When starting an optimization with the VMEC-based approach from either the jorge\_QA or the rCSSC initial guesses, the optimizer consistently tries to enter regions of parameter space where VMEC fails to find an equilibrium, leading either to poor optima, early termination, or configurations without magnetic surfaces. In consequence, all results shown in Figure \ref{fig:vmec_vs_Boozer_surface_configuration_summary} obtained with the VMEC-based approach are obtained from the CNT32 and CNT39 initial guesses. The jorge\_QA and rCSSC initial guesses however performs in general better, as shown in Figure \ref{fig:optimum_summary}. The Boozer surface approach does not need to find an exact solution to Eq.(\ref{eq.Boozer_pde}), as it is solved in the least-squares sense, avoiding entirely this issue faced by the VMEC-based approach.

Third, derivatives of the objective function with respect to the degrees of freedom are evaluated with finite-differences approximation in the VMEC-based approach. Finite differences generate numerical errors, and therefore might stop the optimizer from approaching further the local minimizer. This issue could be circumvented by considering an alternative MHD equilibrium code that provides derivatives, such as the DESC code.\citep{panici_2023,conlin_2023,dudt_2023} In the case of the Boozer surface approach, derivatives are obtained by automatic differentiation, providing an accuracy of the order of the machine precision.

Fourth, the VMEC-based approach optimizes a larger number of degrees of freedom than the Boozer surface approach. Indeed, the VMEC-based approach optimizes the coils and the plasma boundary at the same time, while the Boozer surface approach only optimizes the coils --- the plasma boundary is an output of the calculation. 

Overall, the VMEC-based approach is thus less robust than the Boozer-surface approach, as it requires more fine-tuning from the user, and often fails in finding a good optimum with nested magnetic surfaces. As a side note, the VMEC-based approach is noticeably slower than the Boozer surface approach, at it optimizes more degrees of freedom, and relies on finite-differences to evaluate the objective function gradient. For all these reasons, all results presented thereafter are obtained with the Boozer-surface approach.


Note that the Boozer surface approach comes with its own challenges. For example, it is found to be difficult to target a plasma volume very different from the initial guess volume. This is why the rCSSC initial guess has been constructed to have the targeted plasma volume, at the expense of coils not satisfying the vessel constraint. Another approach, to rescale the initial guess such that the coils fit within the vessel, led to disappointing results precisely because the newly constructed initial guess volume was too small, and the optimizer could not find a path towards a minimizer with the targeted volume.

\begin{figure}
    \centering
    \begin{tikzpicture}
        \node (fig1) at (0,5.5) {\includegraphics[width=.7\linewidth]{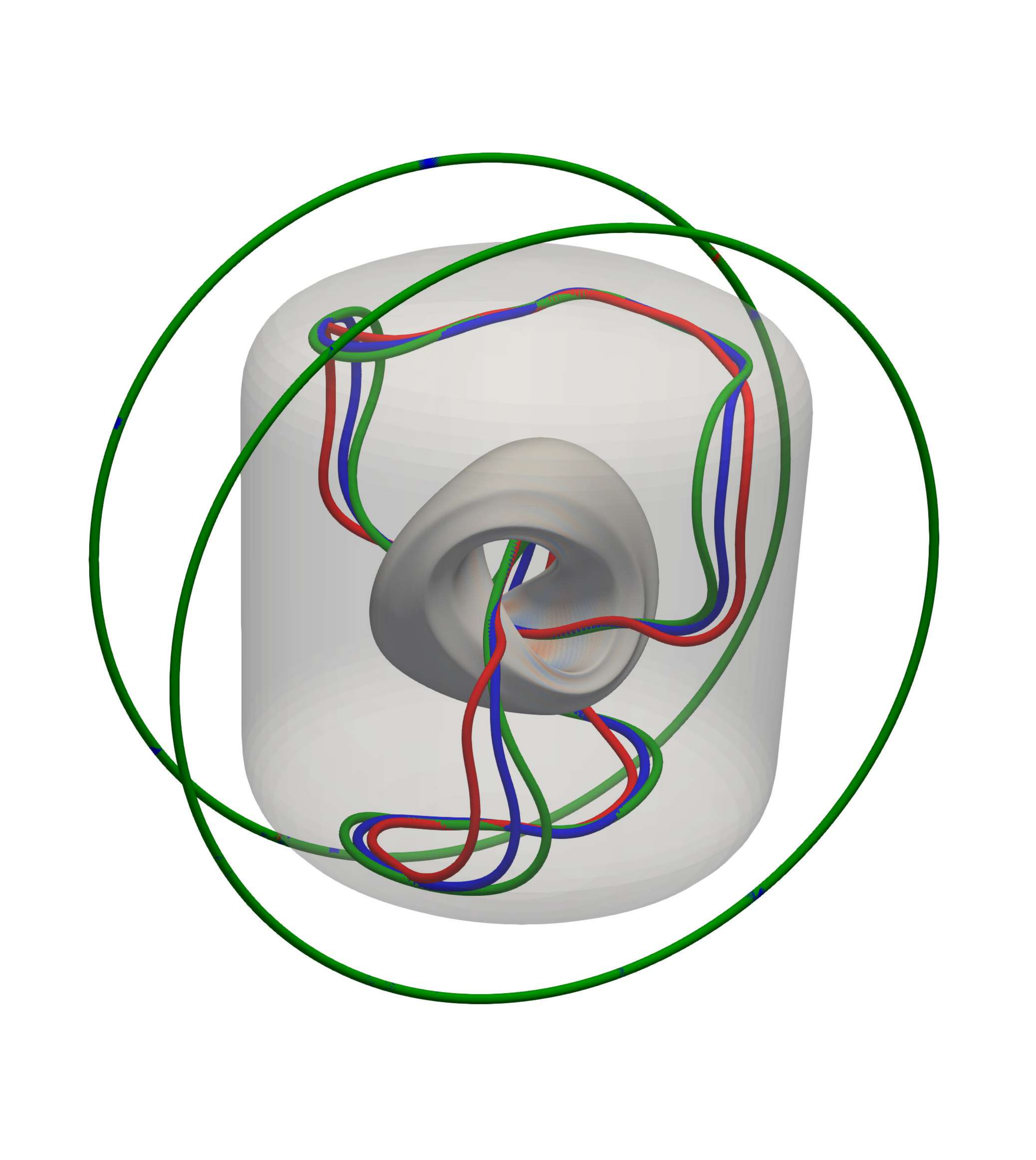}};
        \node (fig2) at (-5,-3) {\includegraphics[height=6cm]{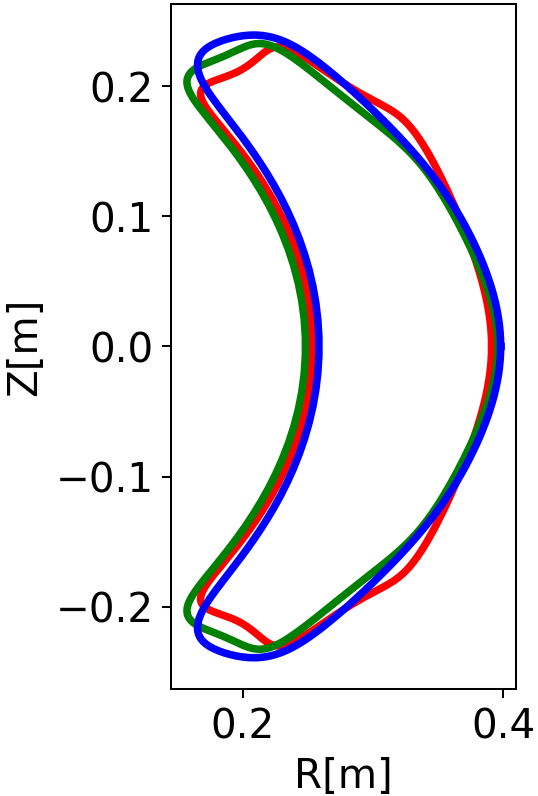}};
        \node (fig2) at (2,-3) {\includegraphics[height=6cm]{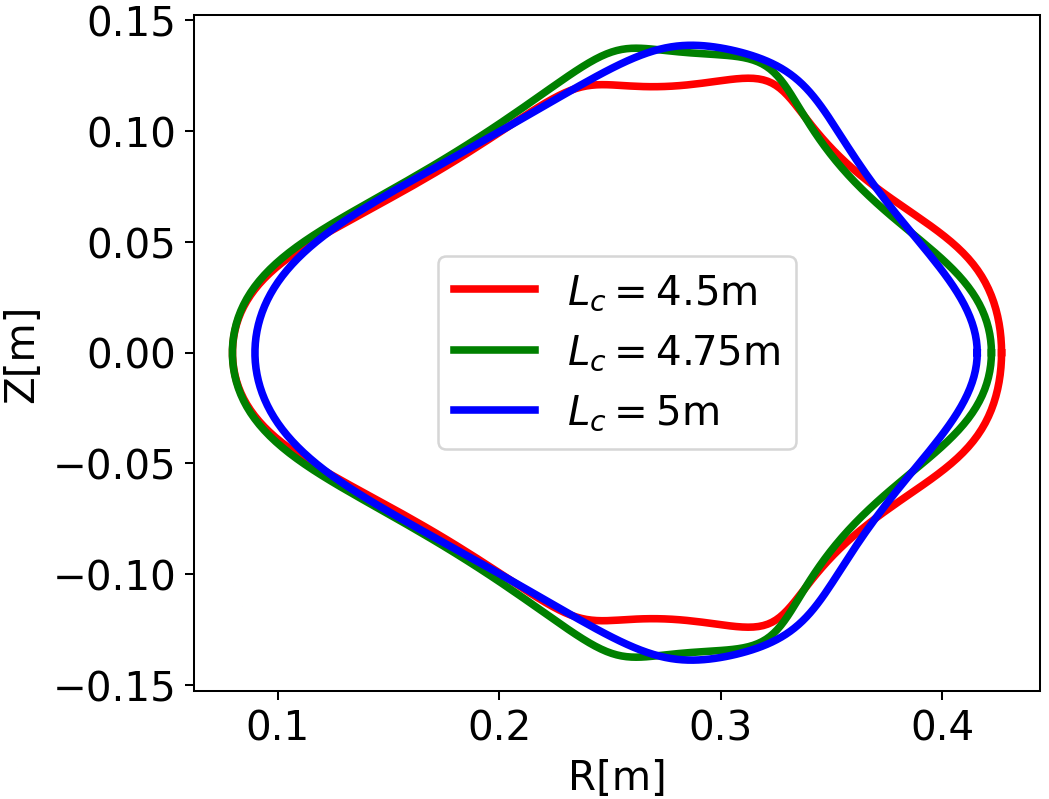}};
    \end{tikzpicture}
    \caption{Top: 3-dimensional plot of three different configurations obtained from the rCSSC initial guess, for IL coils length of $L_c=4.5$m (red), $L_c=4.75$m (blue), and $L_c=5$m (green). Bottom: plasma bean (left) and triangular (right) cross-sections for the three different configurations. These configurations correspond to the circles in Figure \ref{fig:optimum_summary}.}
    \label{fig:coil_length_comparison}
\end{figure}

We now discuss some general features of the optima shown in Figure \ref{fig:vmec_vs_Boozer_surface_configuration_summary} and \ref{fig:optimum_summary}. All configurations have a maximum HTS strain lower than the engineering limit of $0.004$. Note that lowering the engineering threshold, as it has been done to generate the "rCSSC low strain" set of solutions, has only a minimal impact on the QS error. This indicates that minimal deformation of the coils can greatly help to mitigate the strain on the HTS tape. Regarding the coil length, it is no surprise that when the constraint on the coil length is relaxed, solutions with lower QS error are found. For example. see Figure \ref{fig:coil_length_comparison}, where the three configurations obtained from the rCSSC initial guess for different lengths of the IL coils are plotted. We note that the coils and the plasma shape are very similar. The additional coil length is used by the optimizer to further shape the plasma, while staying in the same neighborhood of solution in parameter space.  

We now highlight five configurations selected for further analysis, summarized in Table \ref{tab:highlights}. The \emph{CSX\_LS\_4.5} configuration has the lowest HTS strain among the candidate configurations, and its IL coils are short. The \emph{CSX\_4.75} has better QS at the price of longer coils and larger HTS strain. The \emph{CSX\_5.0\_1} and \emph{CSX\_5.0\_2} are candidates with longer coils, leading to small QS error. Finally, the \emph{CSX\_WPs\_5.0} is a proposed upgrade to CSX with additional coils, as described in section \ref{sec.wps}. 

\begin{figure}
    \centering
    \includegraphics[width=.8\linewidth]{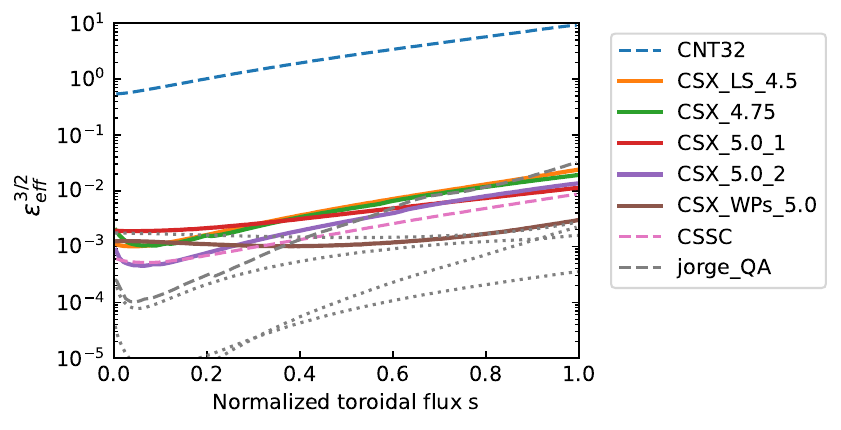}
    \caption{Effective ripple as a function of the normalized toroidal flux for the selected configurations of Table \ref{tab:highlights} (full lines), and some of the considered initial guesses (dashed lines). Other neo-classically optimized experiments (HSX, W7-X, CFQS, and NCSX) are shown in gray, dotted lines for comparison.}
    \label{fig:effective_ripple}
\end{figure}

We evaluate the effective ripple $\epsilon_{eff}$ \citep{lotz_1985,beidler_1995} using the NEO code \citep{nemov_1999a} for each of the selected configurations and compare it to some other common designs in Figure \ref{fig:effective_ripple}. Through optimization, the effective ripple is improved by approximately three orders of magnitude in comparison to the CNT device, ensuring a better neoclassical confinement. Note that the other initial guesses, \textit{i.e.} the CSSC and jorge\_QA configurations, have an effective ripple of the same order as the CSX configurations, demonstrating that the optimization was able to find a solution that satisfies the engineering constraints without affecting the initial guesses good confinement properties. 

\begin{table}
    \setlength{\tabcolsep}{10pt}
    \centering
    \begin{tabular}{l|rrrrr}
      \textbf{Name}  & \textbf{Initial guess} & $\mathbf{L_c}$ & $\mathbf{f^b_{QA}(\Gamma_b)}$  & $\mathbf{max(\boldsymbol{\epsilon}_b+\boldsymbol{\epsilon}_t)}$  \\
      \hline\hline
      CSX\_LS\_4.5   & rCSSC   & 4.5m & 5.71\%  & $2.2\cdot 10^{-3}$\\
      CSX\_4.75   & CNT32  &  4.75m  & 5.45\%  & $3.5\cdot 10^{-3}$\\
      CSX\_5.0\_1   & rCSSC  & 5.0m  & 5.10\%  & $2.5\cdot 10^{-3}$\\
      CSX\_5.0\_2   & jorge\_QA  & 5.0m  & 4.98\%  & $3.7\cdot 10^{-3}$\\
      \hline
      CSX\_WPs\_5.0   & rCSSC  & 5.0m  & 3.08\%  & $2.1\cdot 10^{-3}$\\
      \hline
    \end{tabular}
    \caption{Selected configurations for CSX}
    \label{tab:highlights}
\end{table}

\begin{figure}
    \centering
    \includegraphics[width=0.475\linewidth]{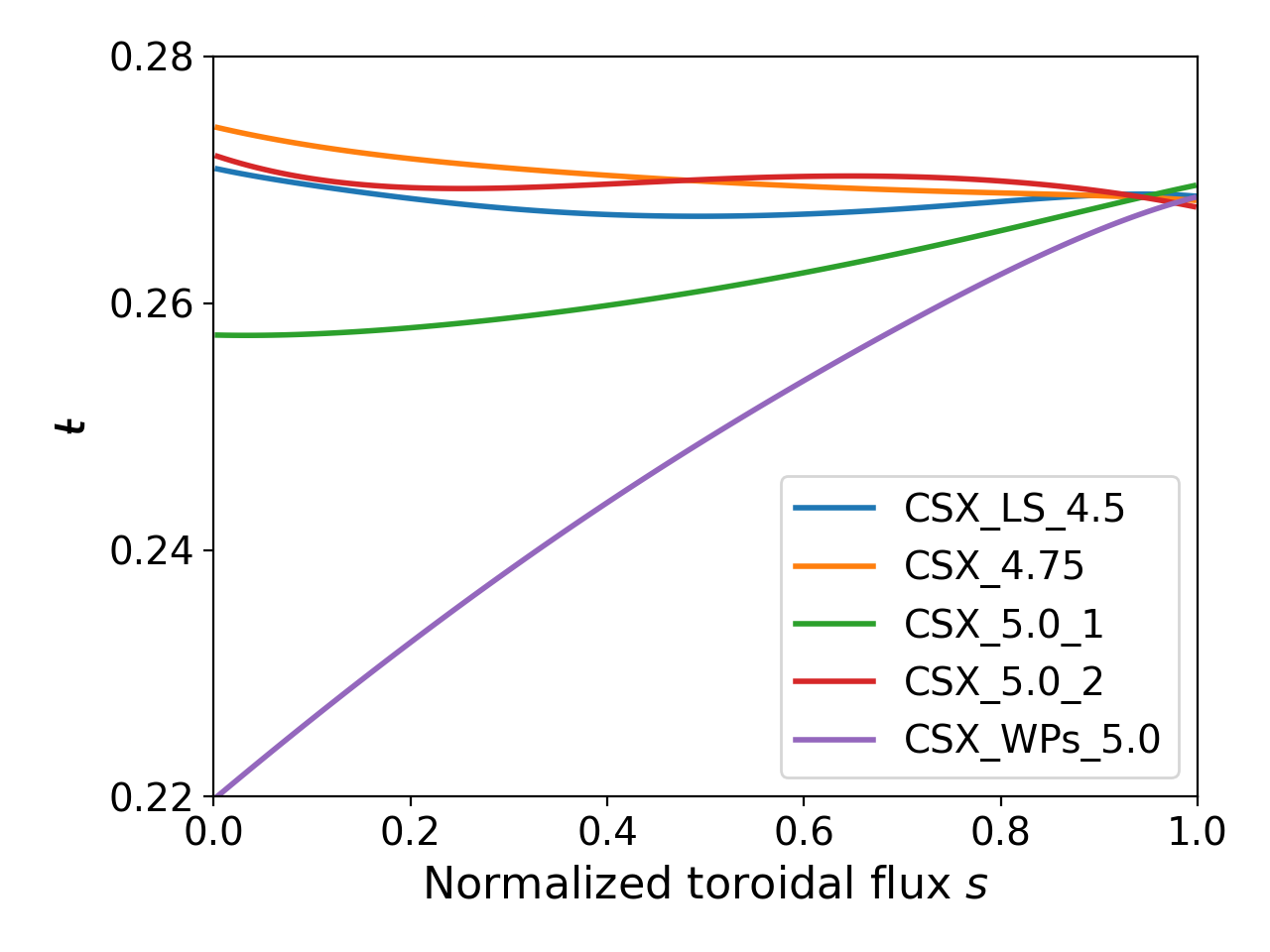}
    \hfill
    \includegraphics[width=0.475\linewidth]{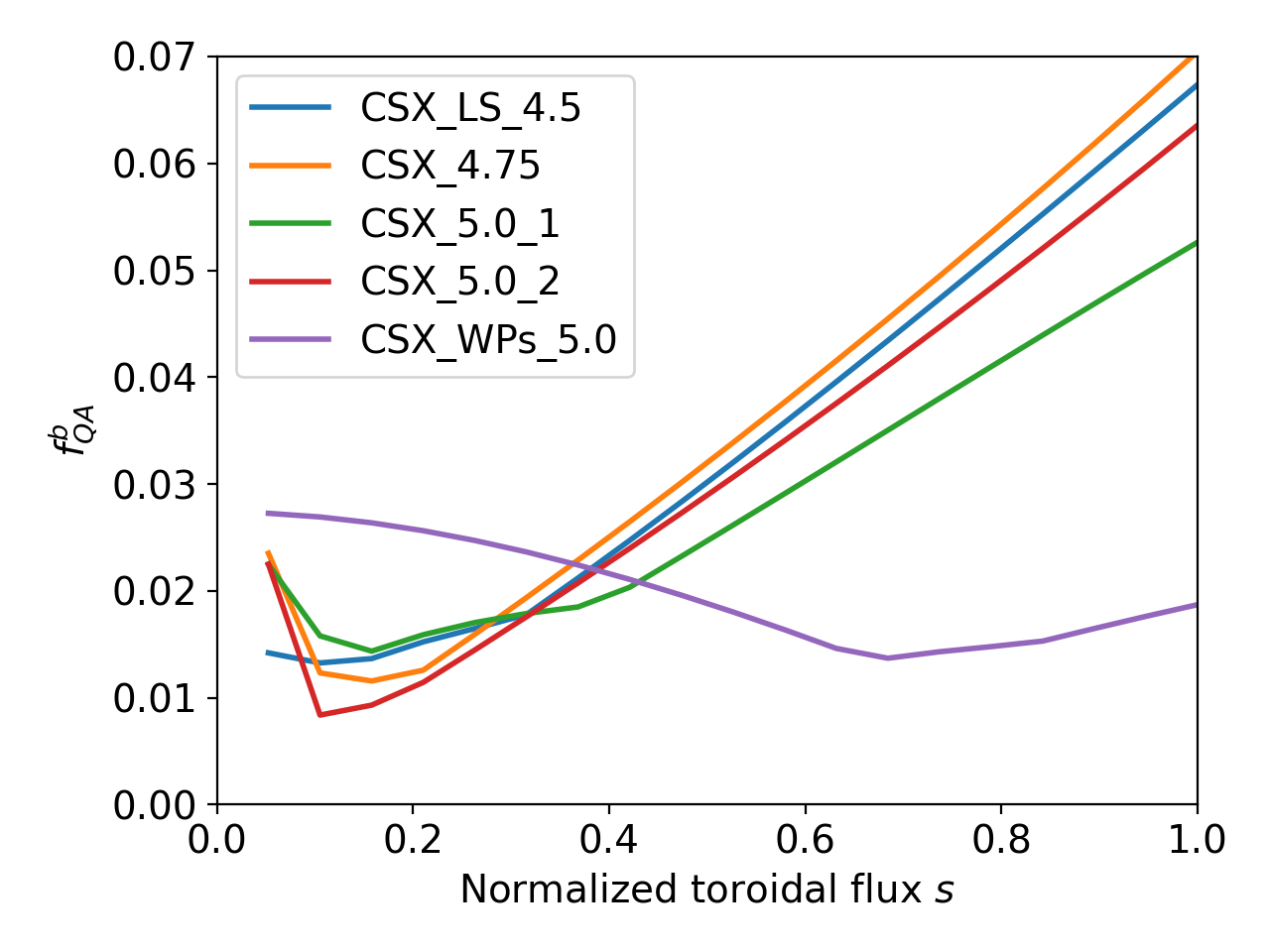}
    \caption{Rotational transform (left) and QS error (right) as a function of the normalized toroidal flux for the selected configurations of Table \ref{tab:highlights}.}
    \label{fig:profiles}
\end{figure}

The rotational transform profile and QS error are plotted in Figure \ref{fig:profiles} for each of the selected configurations. The QS error is mostly below the threshold value of $5-10\%$ required to sustain plasma flow, as discussed in section \ref{sec:plasma objectives}. We observe that, from a plasma physics point of view, all devices are similar to one-another, except for the \emph{CSX\_WPs\_5.0}, which will be described in section \ref{sec.wps}. Choosing one design among these candidates therefore requires a careful analysis of their engineering properties. Further studies will be conducted, for example on the configuration sensitivity to coil displacement and deformation,\citep{landreman_2018a,wechsung_2022a} on coil forces \citep{hurwitz_2023}, and on considering the effect of coils with finite width.\citep{singh_2020,mcgreivy_2021} In addition, coils prototypes are currently being tested to gain further insight on the IL coils manufacturing.

\begin{figure}
    \centering
    \hfill
    \includegraphics[height=7cm]{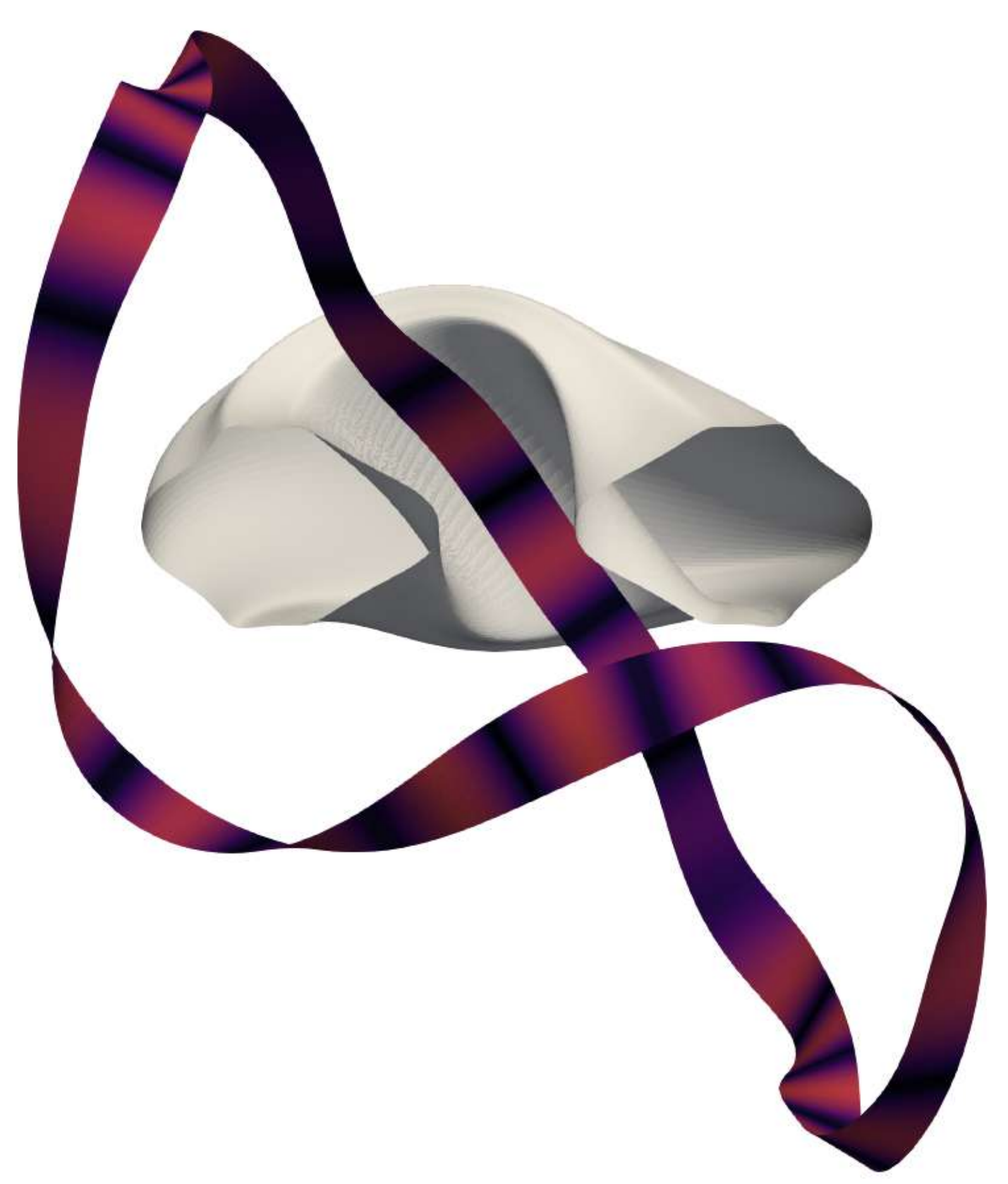}
    \hfill
    \includegraphics[height=6cm]{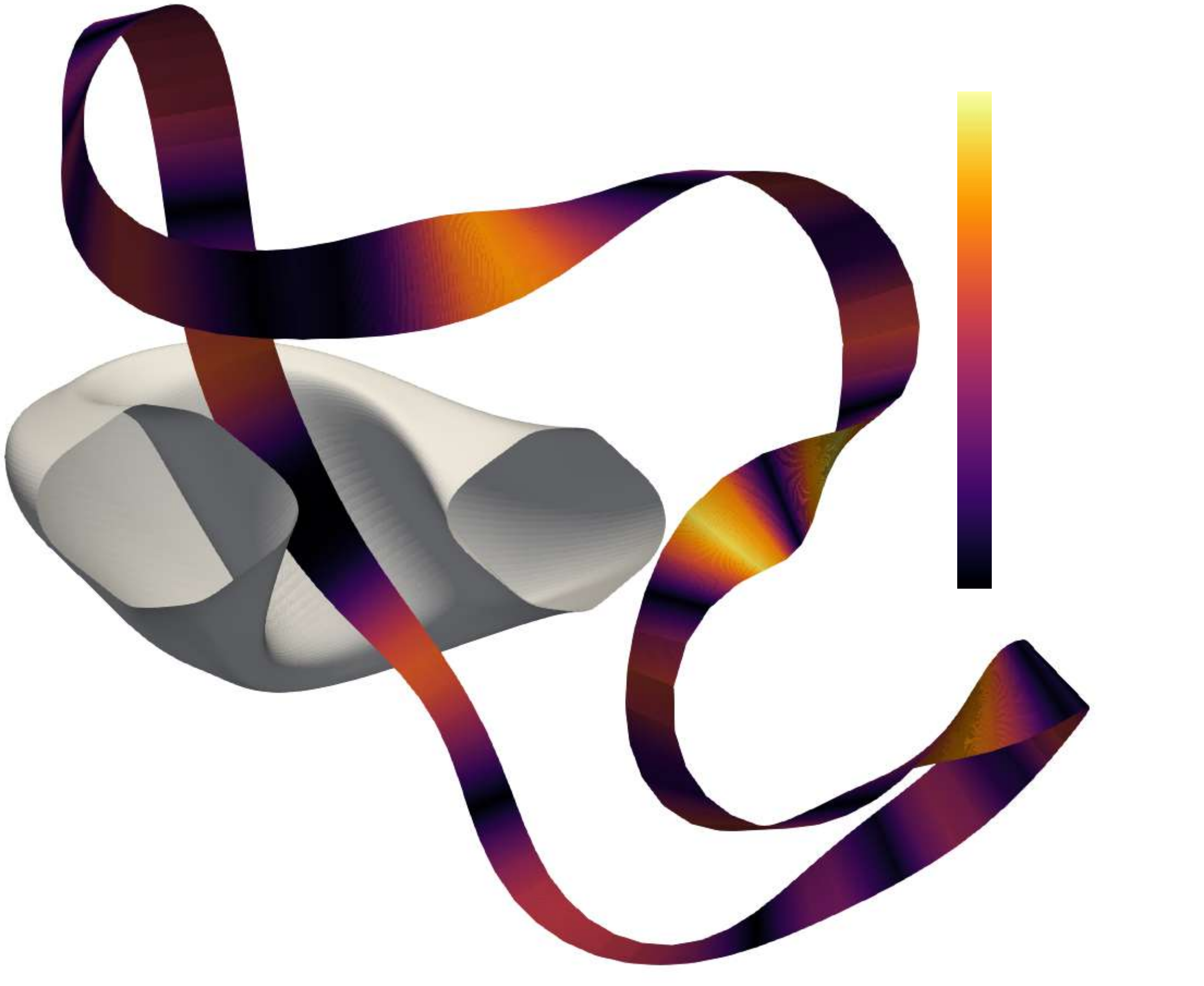}
    \hfill
    \caption{3D plot of the HTS tape frame for the \emph{CSX\_LS\_4.5} (left) and \emph{CSX\_5.0\_2} (right) configurations. Colors indicate the local strain on the HTS tape. The tape width is here enhanced for visualization purposes.}
    \label{fig:hts_frame}
\end{figure}

In Figure \ref{fig:hts_frame} the HTS winding frame is plotted for the \emph{CSX\_LS\_4.5} configuration, which has the lowest HTS strain of the selected configurations, and for the \emph{CSX\_5.0\_2} configuration, which has the highest. Interestingly, we notice that in the case of the \emph{CSX\_5.0\_2} configuration, the largest HTS strain occurs the furthest away from the plasma. Slight modifications of the coils could potentially reduce the HTS strain on these critical points without strong impact on the plasma.

\begin{figure}
    \centering
    \begin{tikzpicture}
        \node (fig1) at (-5.5,0) {\includegraphics[height=12cm]{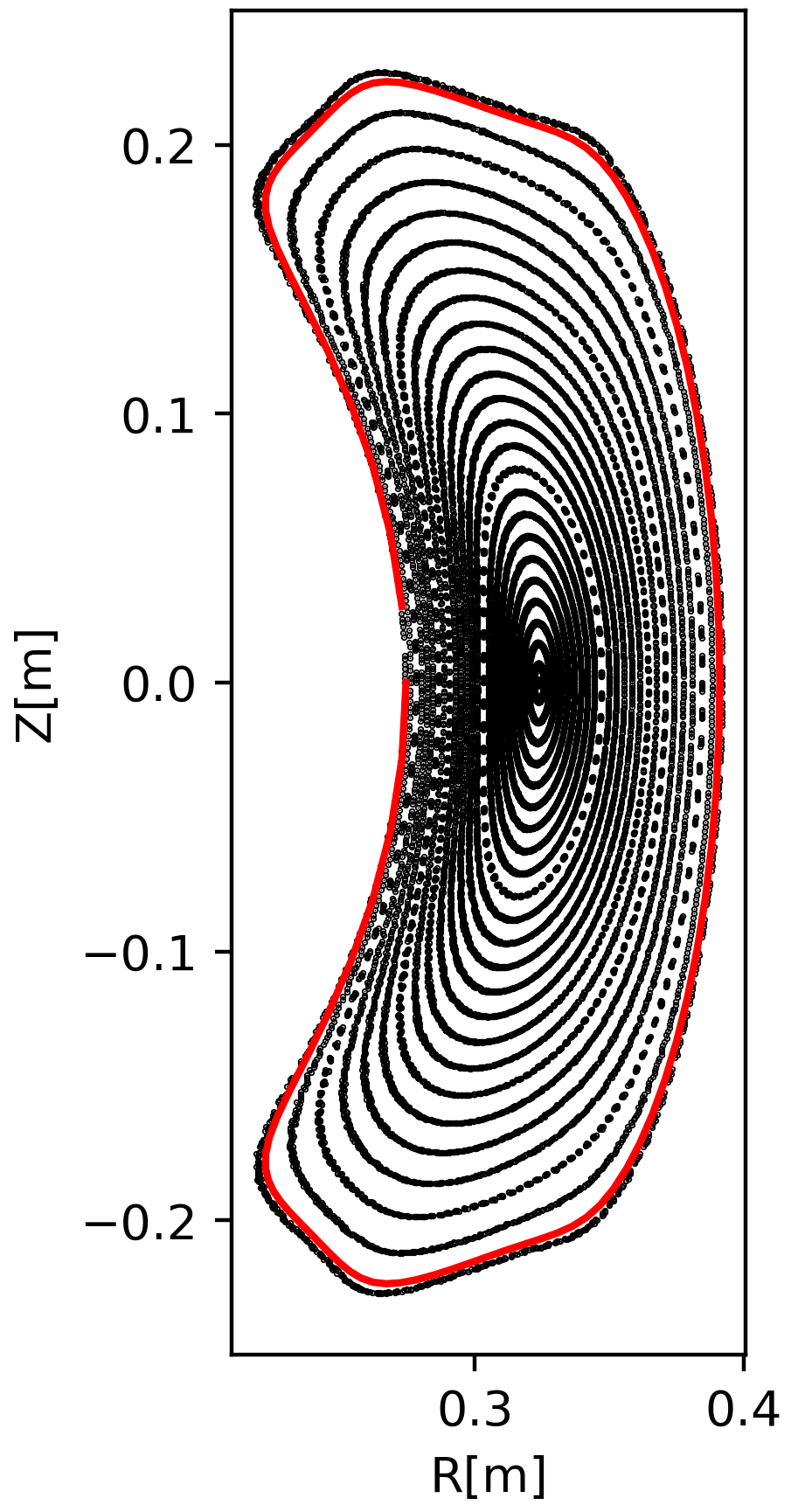}};
        \node (fig2) at (2.6,3.1) {\includegraphics[width=.45\linewidth]{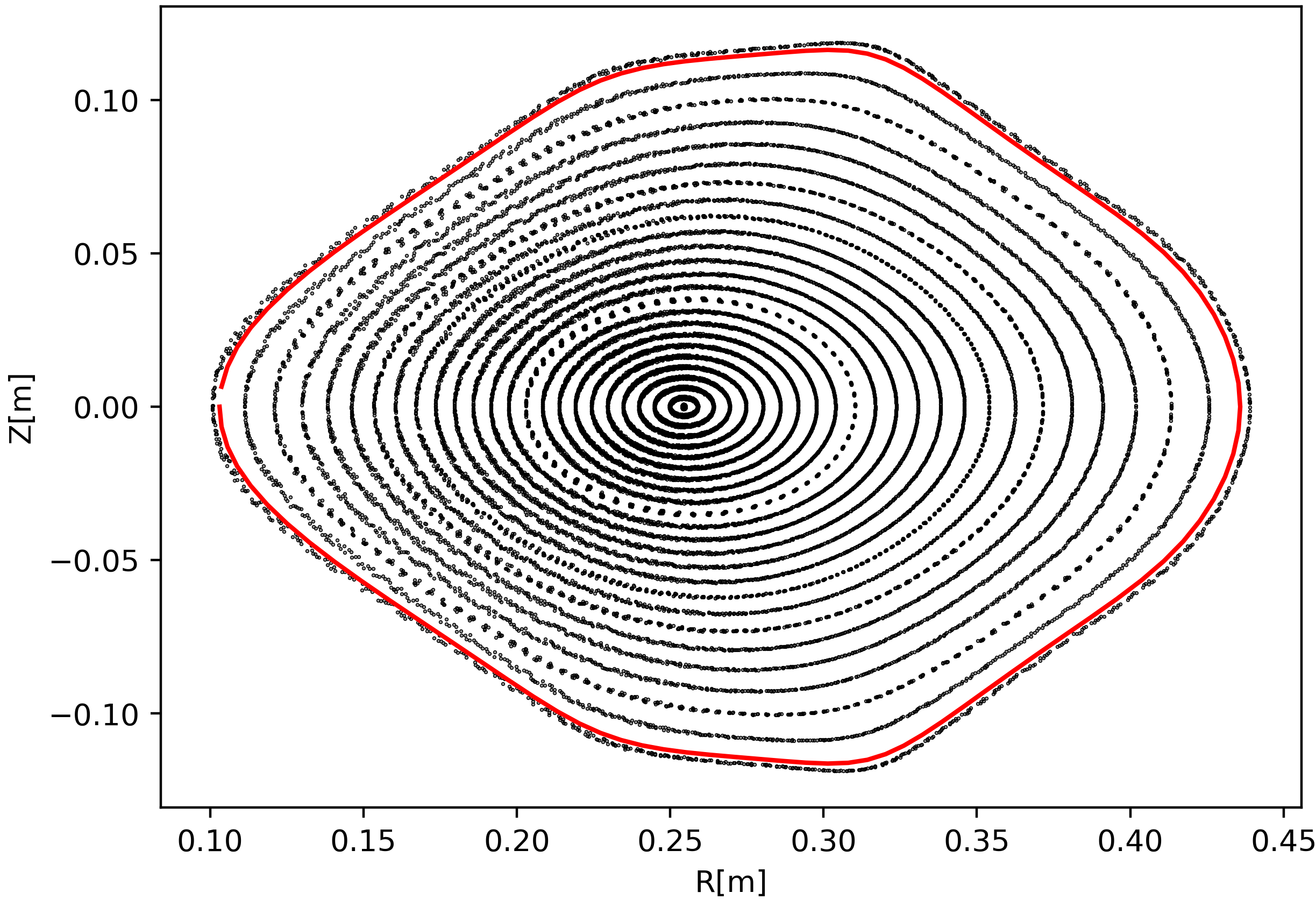}};
        \node (fig2) at (3,-2.5) {\includegraphics[width=.5\linewidth]{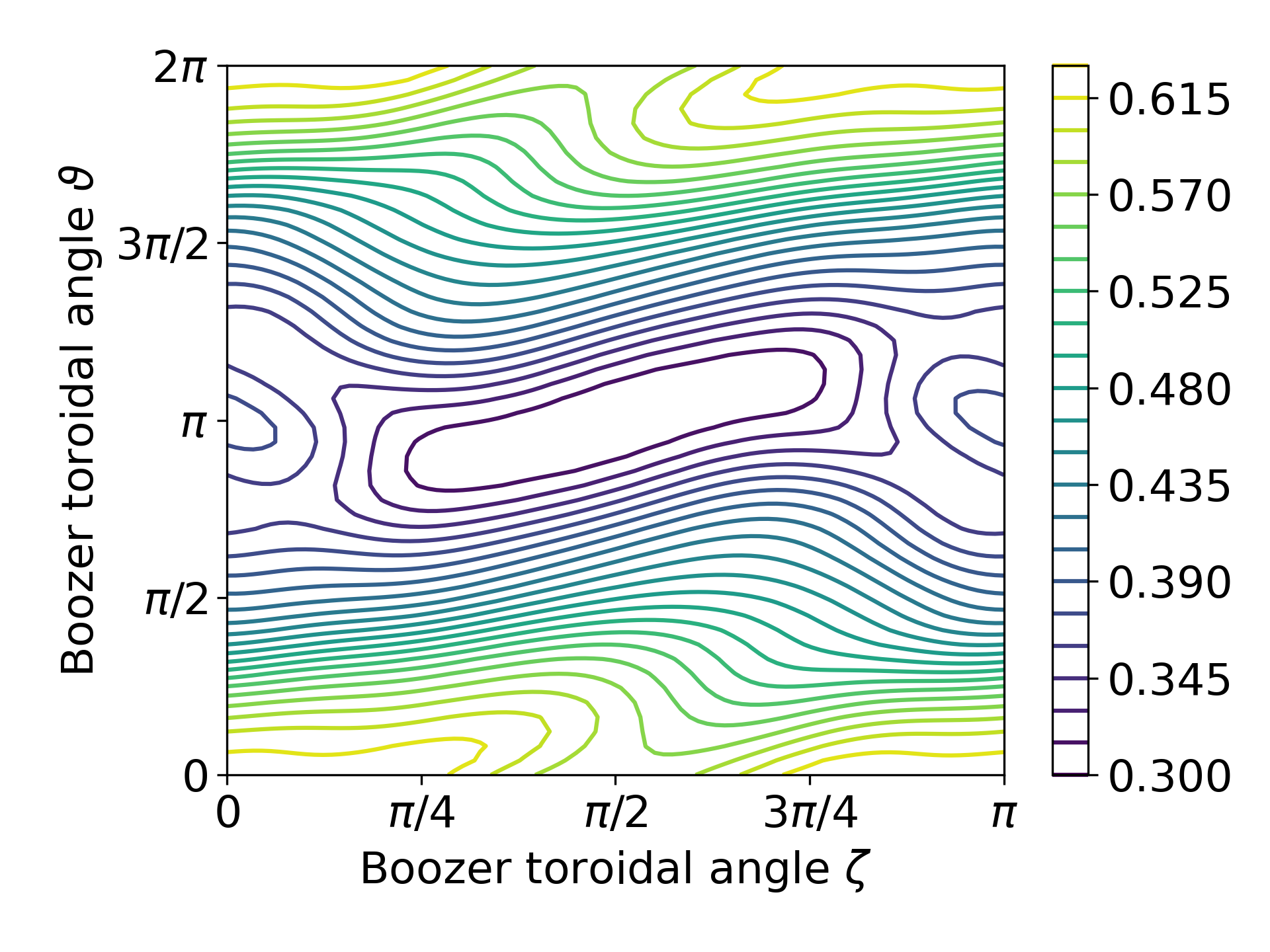}};
    \end{tikzpicture}
    \caption{Bean-shaped (left) and triangular (top right) Poincar\'e section of configuration \emph{CSX\_5.0\_2}. The red surface is $\Gamma_b$, \textit{i.e.} the surface used to evaluate $f^{Boozer}_I$. Bottom right: constant $|B|$ lines on $\Gamma_b$ for the \emph{CSX\_5.0\_2} configuration. Colors indicate $B$ in $[T]$. }
    \label{fig:csx1_poincare}
\end{figure}

The Poincar\'e section of the \emph{CSX\_5.0\_2} configuration is shown in Figure \ref{fig:csx1_poincare}. The configurations \emph{CSX\_LS\_4.5}, \emph{CSX\_4.75}, and \emph{CSX\_5.0\_1} have similar Poincar\'e sections and are not shown here. The contours of constant field strength on $\Gamma_b$ in Boozer coordinates are shown on the bottom right panel of Figure \ref{fig:csx1_poincare}.

\subsection{Improving quasi-axisymmetry by including additional windowpane coils} \label{sec.wps}
Configurations presented in section \ref{sec:best_results} satisfy all physics objectives and engineering constraints. Nevertheless, one may wonder in what ways the QS could be improved. We propose here a potential future upgrade to the CSX design, where additional windowpane (WP) coils are included to further shape the plasma.


Mechanical support to these coils is provided by the cylindrical vacuum vessel; the coils are therefore constrained to wrap around the vacuum chamber. We represent the WP coils in cylindrical coordinates $(r,\phi,z)$,  constrain them to have the same radius as the vacuum chamber, $r=R_v$, and to be ellipses in the RZ-plane, with
\begin{align}
    \phi(l) &= \phi_0 + \phi^c_1\cos(2\pi l) + \phi^s_1\sin(2\pi l)\\
    z(l) &= z_0 + z^c_1\cos(2\pi l) + z^s_1\sin(2\pi l),
\end{align}
with $\{\phi_0,z_0,\phi^c_1,\phi^s_1,z^c_1,z^s_1\}$ the degrees of freedom of the curve. As the vacuum vessel is far from the plasma, the WP coils tend to have large currents. We include an additional penalty term to limit the maximum current in the WP coils in Eq.(\ref{eq.coil_regularization}). We write
\begin{equation}
    f_{reg} = \ldots + w_{wps} \sum_{k=1}^{N_{wp}}\max(I_k-I_{max},0)^2,
\end{equation}
where $N_{wp}$ is the number of WP coils, $I_k$ is the current in the $k^{\text{th}}$ WP coil, and $I_{max}$ is the maximum allowed current in the WP coils. Here we set $I_{max}=50\ \text{kA}$, about half the current in the IL coils.

\begin{figure}
    \centering
    \includegraphics[width=0.75\linewidth]{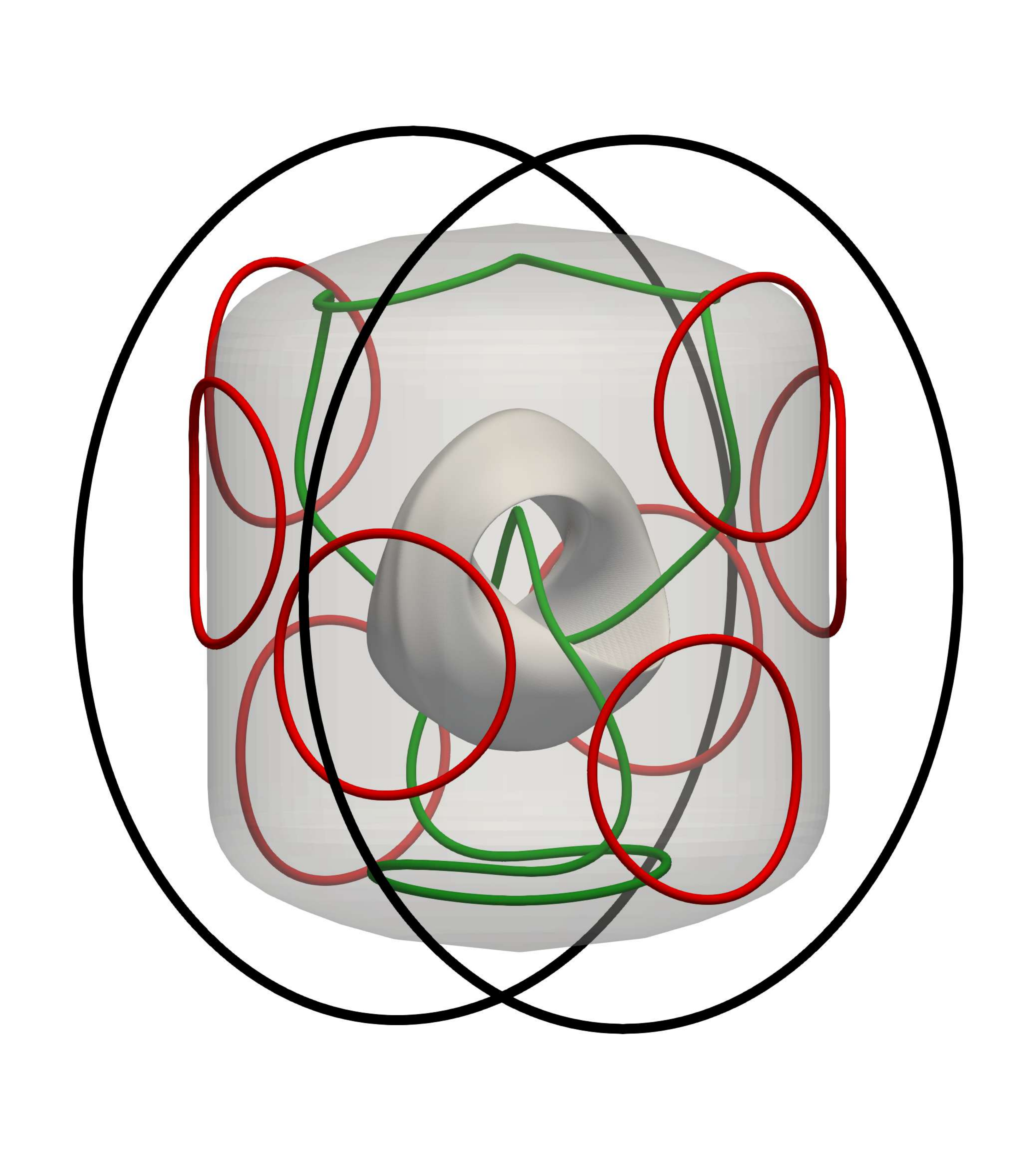}
    \caption{3-dimensional rendering of the \emph{CSX\_WPs\_5.0} configuration. Different coil topologies are plotted in green (IL coils), in red (WP coils), and in black (PF coils)}
    \label{fig:paraview_csx_wps}
\end{figure}
\begin{figure}
    \centering
    \begin{tikzpicture}
        \node (fig1) at (-5.5,.2) {\includegraphics[height=12cm]{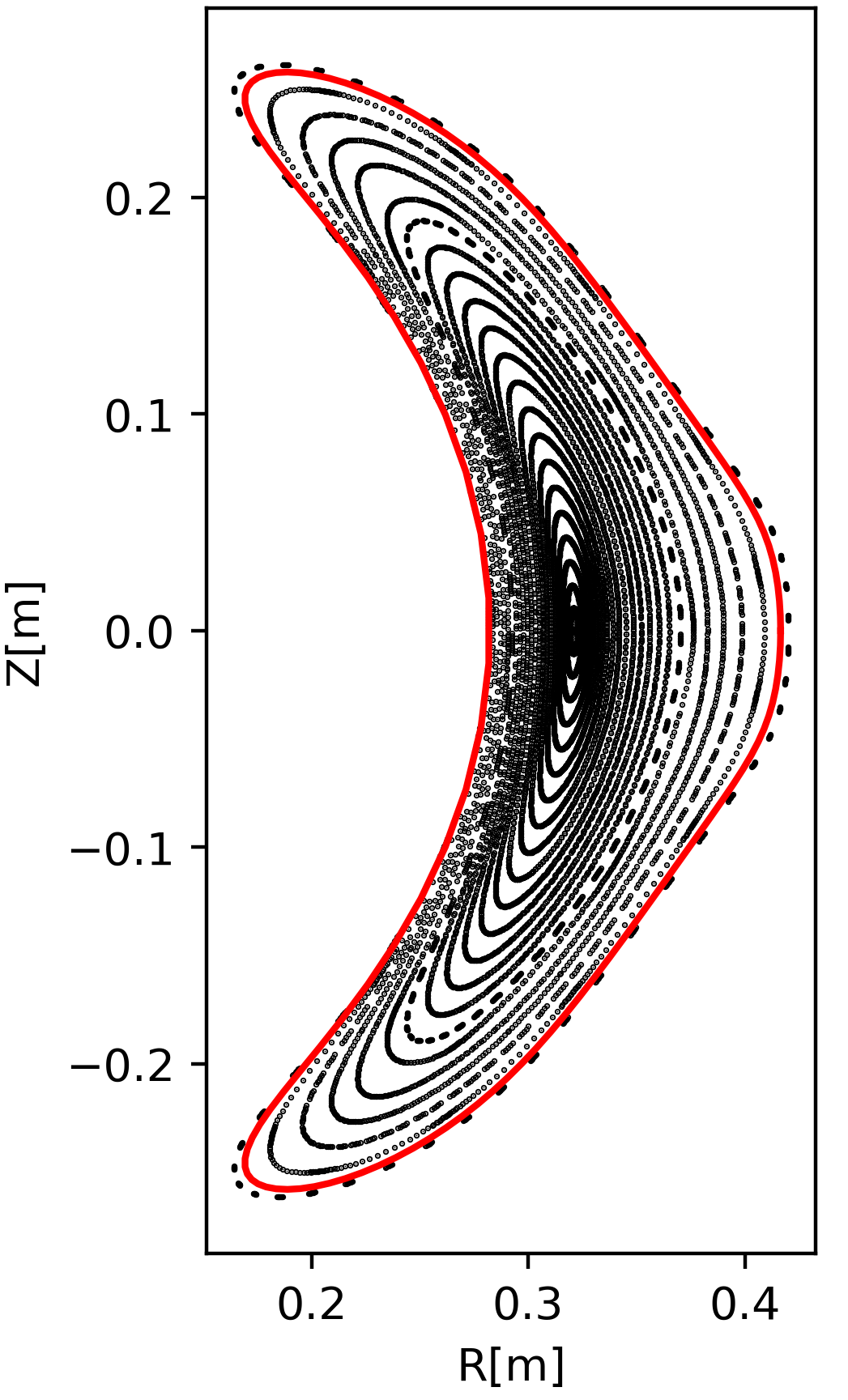}};
        \node (fig2) at (2.4,3) {\includegraphics[width=.375\linewidth]{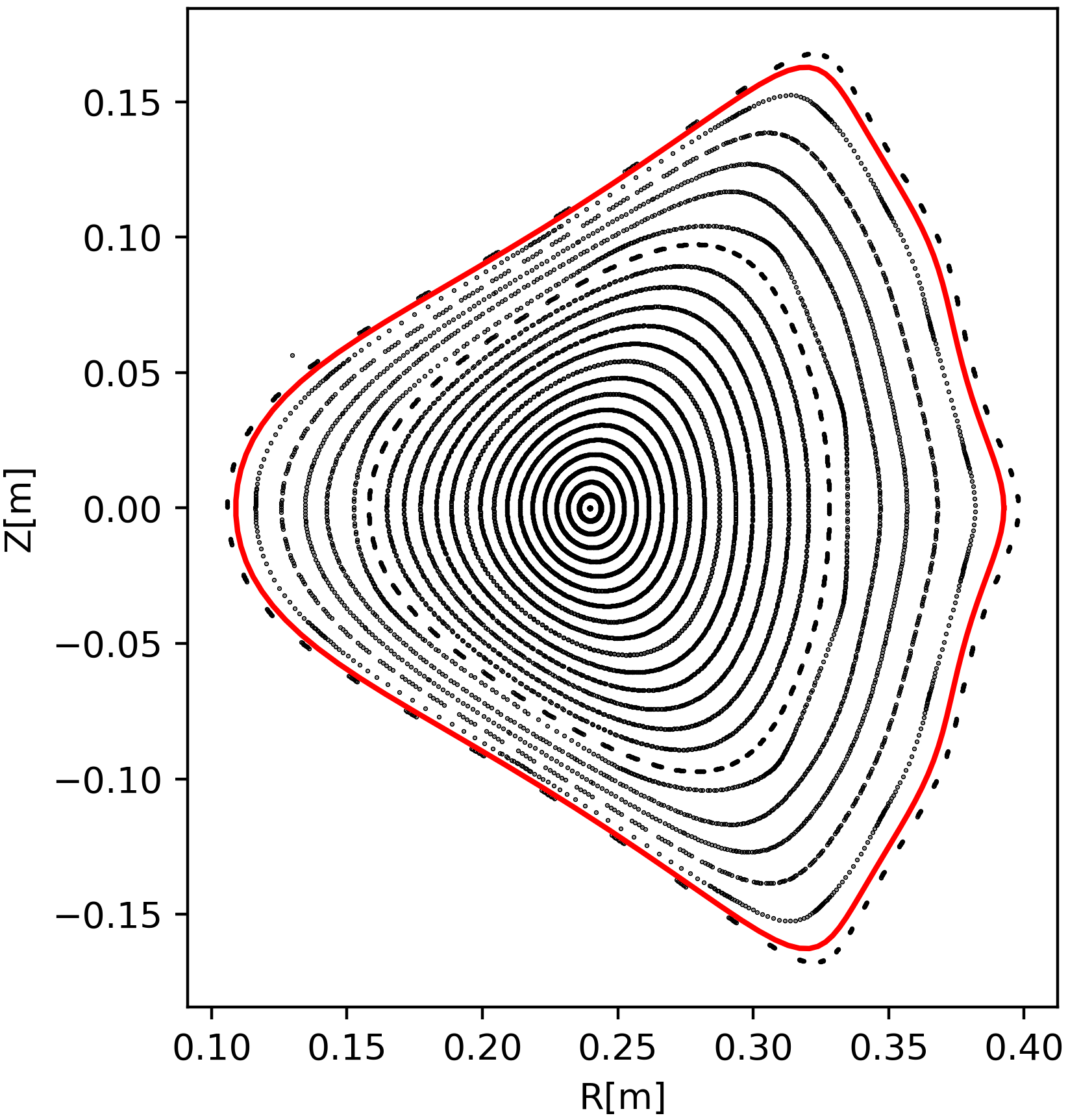}};
        \node (fig2) at (3,-3.5) {\includegraphics[width=.5\linewidth]{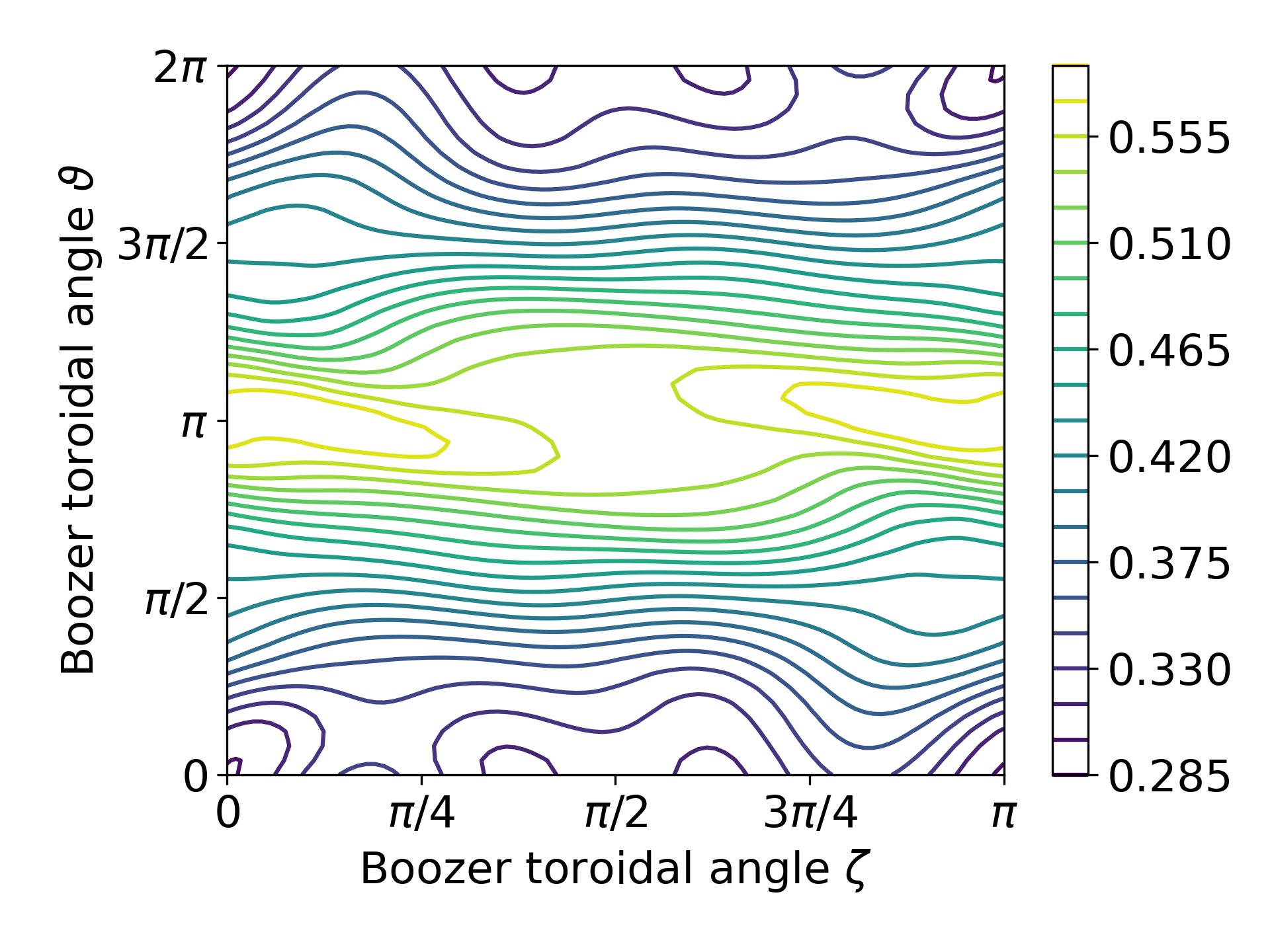}};
    \end{tikzpicture}
    \caption{Bean-shaped (left) and triangular (top right) Poincar\'e section of configuration \emph{CSX\_WPs\_5.0}. The red surface is $\Gamma_b$, \textit{i.e.} the surface used to evaluate $f^{Boozer}_I$. Bottom right: constant $|B|$ lines on $\Gamma_b$ for the \emph{CSX\_5.0\_2} configuration. Colors indicate $B$ in $[T]$. }
    \label{fig.poincare_with_wps}
\end{figure}

The obtained configuration is shown in Figure \ref{fig:paraview_csx_wps}. The rotational transform and QS error profiles are plotted on the left and right of Figure \ref{fig:profiles} respectively, and the effective ripple is plotted in Figure \ref{fig:effective_ripple}. As one would expect, increasing the dimensionality of parameter space by increasing the number of coils leads to better results; the QS error is smaller than any of the other selected configuration for CSX, and the effective ripple is about half an order of magnitude better at the plasma edge. The rotational transform profile however crosses some potentially problematic low-order rationals, in particular the $\iotabar=0.25$ resonance, which might cause the emergence of large magnetic islands. The Poincar\'e section of the \emph{CSX\_WPs\_5.0} configuration is shown in Figure \ref{fig.poincare_with_wps}. Note that despite crossing the resonance $\iotabar=0.25$, no large islands are visible in the plasma. Figure \ref{fig.poincare_with_wps} bottom right panel shows contours of constant field strength on $\Gamma_b$.

Note that even though QS can be improved by using additional WP coils, configurations including WP coils are as-of-now not considered for CSX. While these configurations are closer to QS, the objective of finding a field with QS error below $5-10\%$ is achieved even without using WP coils. Instead, future studies will focus on the possibility to optimize WP coils for island width control, error field correction, and device flexibility.

There are, of course, other ways to increase the number of degrees of freedom of CSX. One example would be to allow the translation of the PF coils along their symmetry axis; this however would require to redesign their stands and carefully reposition them, thereby increasing the complexity of the project and the risk of positioning errors. One might think of dropping the constraint of stellarator symmetry, roughly doubling the number of degrees of freedom; non-stellarator symmetric plasmas are however generally less studied than stellarator symmetric ones. The analysis of a non-stellerator symmetric plasma would be challenging, as many numerical codes assume the plasma to be stellarator symmetric. One final way to increase the number of degrees of freedom in the optimization would be to consider a single-field period device, and therefore design two different IL coils. These ideas are, in principle, feasible from an engineering point of view, though they complicate the design and construction of the device. Further studies are required to determine if any of these proposals could bring sufficient plasma improvements when compared to the additional cost and engineering challenges.



\section{Conclusion}
This paper discussed the plasma objectives and engineering constraints of the CSX experiment currently being designed at Columbia University. The experimental goals of CSX are to (i) demonstrate the use of NI-HTS to design 3-dimensional stellarator coils, and (ii) test some of the theoretical predictions of neoclassical theory in QA magnetic fields, in particular measuring the plasma flow damping in the direction of symmetry. These experimental goals were translated into target plasma parameters, such as a maximum QA breaking mode of $5$ to $10\%$, a minimum rotational transform of $0.27$, and a plasma volume of $0.1\ \text{m}^3$ filled with magnetic surfaces.

The CSX experiment will refurnish most of the equipment of the CNT experiment, and replace the IL coils with optimized coils to achieve QA. Multiple engineering constraints have to be satisfied, in particular geometric constraint, \textit{i.e.} the coils should fit within the existing vacuum vessel, not intersect other coils, and should remain sufficiently far from the plasma. In addition, the strain on the HTS tape should remain below a critical value. The challenge to find an optimized configuration that matches the physics objectives while satisfying the engineering constraints is only achieved by considering combined plasma-coil optimization algorithms, where both the plasma shape, the coils shape, and the HTS tape orientation are optimized at the same time.


Two single-stage optimization algorithms were used, namely the \emph{VMEC-based approach} \citep{jorge_2023} and the \emph{Boozer surface approach}.\citep{giuliani_2022,giuliani_2022a,giuliani_2023} While the VMEC-based approach found multiple interesting optima, they were in general further from QA than the configurations found with the Boozer surface approach --- see Figure \ref{fig:vmec_vs_Boozer_surface_configuration_summary}. In addition, it was found that the VMEC-based approach required a time-consuming manual fine-tuning of numerical weights; failure to do so often lead to configurations without magnetic surfaces. The Boozer surface approach, on the other hand, was found to be computationally less expensive, more robust, and easier to use than the VMEC-based approach. Overall, the Boozer surface approach was therefore more successful when applied to the specific problem of the CSX optimization. 

Among all configurations found in this paper, four were selected, with the addition of one exotic configuration in which windowpane coils were included to further shape the plasma --- see Table \ref{tab:highlights}. All selected configuration shared similar values of QA, and similar rotational transform profile, indicating that all optima were close in parameter space. Deciding which configuration should be built for the CSX experiment requires further work. In particular, further focus will be on the sensitivity of each configuration to coils manufacturing errors and coils deformation, and on the effect of finite-width conductors on the magnetic field. Coils prototypes, wound with steel tape, will provide additional insights on the winding difficulties of each configuration.

\newpage
\section*{Acknowledgements}
The authors would like to thank the CSX engineering team, the EPOS team, S. R. Hudson, M. C. Zarnstorff, A. Giuliani, A. Kaptanoglu, D. Biek, R. Jorge and the MUSE team for useful discussion.

\section*{Conflict of interest}
The authors report no conflict of interest.

\appendix

\section{Rotational transform constraint} \label{app.iota_constraint}
Over an orbit, a trapped particle will drift across flux surfaces given by \citep{helander_2014}
\begin{equation}
    \Delta\psi_t = \frac{2\pi G}{\iotabar}\frac{m\Delta v_\parallel}{ZeB}, \label{eq.banana_width_helander}
\end{equation}
with $m$ the particle mass, $Z$ the ion charge number, $e$ the elementary charge, and $2\pi G$ the total poloidal current outside the plasma. Here we split the particle velocity in its component parallel and perpendicular to the magnetic field, $\mathbf{v}_0 = v_\parallel \hat{\mathbf{b}} + \mathbf{v}_\perp$, with $\hat{\mathbf{b}}=\mathbf{B}/B$, and $v_0$ the particle velocity, that we estimate to be the particle thermal velocity. The largest difference in parallel velocity, $\Delta v_\parallel$, will be experienced by barely trapped particles. We consider two points along a banana orbit of a barely trapped particle, with the first point where the magnetic field is at its minimum, $B_1=B_{min}$, while the second point is at the particle bounce point, where the particle experience a maximum magnetic field, $B_2=B_{max}$. By conservation of the magnetic moment $\mu=mv_\perp^2/2B$, we write
\begin{equation}
    v_{\perp,2}^2 B_{min} = v_{\perp,1}^2 B_{max},
\end{equation}
with $v_{\perp,i}$ the perpendicular velocity at position $i$. We now estimate $B_{max}\sim B_0(1+\epsilon)$ and $B_{min}=B_0(1-\epsilon)$, with $B_0$ the magnetic field on axis and $\epsilon=r/R_0$ the inverse aspect ratio, and obtain
\begin{equation}
    v_{\parallel,1}^2 = v_0^2 - v_{\perp,1}^2 = v_0^2\left(\frac{2\epsilon}{1+\epsilon}\right),
\end{equation}
where we used $v_{\perp,2}=v_0$, \textit{i.e.} the particle has a purely perpendicular velocity at the bounce point. The difference in parallel velocity is then $\Delta v_\parallel = 2v_{\parallel,1}=2v_0\sqrt{2\epsilon/(1+\epsilon)}$, which leads to
\begin{equation}
    \Delta r \sim \frac{2\rho}{\iotabar}\sqrt{\frac{2}{\epsilon(1+\epsilon)}},
\end{equation}
where we approximated $\psi_t\sim\pi r^2B_0$ in Eq.(\ref{eq.banana_width_helander}), and $G\sim \epsilon a B_0$. Here $\rho_i=mv_0/ZeB$ is the ion gyro radius. We now constrain the rotational transform to be sufficiently large such that the banana orbit width is less than one third of the device minor radius $a$,
\begin{equation}
    \iotabar > \iotabar_c = \frac{6\rho}{a}\sqrt{\frac{2}{\epsilon(1+\epsilon)}}. \label{eq.iota_constraint}
\end{equation}
For the expected plasma parameters of $B_0\sim 0.1T$, $R_0\sim0.27m$, $a\sim0.13m$, and $T_i\sim 5eV$, we obtain $\iotabar_c=0.25$. In this work, we target a mean rotational transform of $\bar{\iotabar}\sim 0.27$. This number has been chosen as it satisfies Eq.(\ref{eq.iota_constraint}), and it is not close to a low-order rational number. This is a desirable property, as if $\iotabar = p/q$ with $q$ small, large magnetic islands can form.

\bibliography{biblio}

\end{document}